\documentclass{pinchcrn}

\usepackage{amsmath,amsfonts,braket,slashed,amssymb,tikz,bm,psfrag,graphicx,color,dsfont}
\newcommand{\pslash}{\rlap{\hspace{0.25mm}/}{p}}

\title{Renormalization Theory\\ and Effective Field Theories}

\author{Matthias Neubert}
\affiliation{PRISMA Cluster of Excellence \& Mainz Institute for Theoretical Physics\\
Johannes Gutenberg University, 55099 Mainz, Germany\\[1mm]
and\\[1mm]
Department of Physics \& LEPP, Cornell University, Ithaca, NY 14853, U.S.A.}

\begin{document}

\maketitle

\preface

These lectures review the formalism of renormalization in quantum field theories with special regard to effective quantum field theories. It is well known that quantum field theories are plagued by ultraviolet (UV) divergences at very short distance scales and infrared (IR) divergences at long distances. Renormalization theory provides a systematic way in which to deal with the UV divergences. Effective field theories deal with the separation of physics on different length or energy scales. The short-distance physics is described by means of Wilson coefficient functions, whereas the long-distance physics is contained in the matrix elements of effective operators built out of the quantum fields for the low-energy effective degrees of freedom of the theory. Renormalization theory is as important for effective field theories as for conventional quantum field theories. Moreover, building on the Wilsonian approach to renormalization, effective field theories provide a framework for a deeper understanding of the physical meaning of renormalization. While the subject of renormalization theory is treated in every textbook on quantum field theory (see e.g.\ \cite{Itzykson:1980rh,Pascual:1984zb,Peskin:1995ev,Collins:2011zzd,Weinberg:1995mt,Weinberg:1996kr,Schwartz:2013pla}), more advanced topics such as the renormalization of composite operators, the mixing of such operators under scale evolution and the resummation of large logarithms of scale ratios are not always treated in as much detail as they deserve. Because of the central importance of this subject to the construction of effective field theories, this course summarizes the main concepts and applications in a concise manner. This course thus sets the basis for many of the more specialized lecture courses delivered at this school. 

These notes assume that the reader has taken an in-depth course on quantum field theory at the graduate level, including some exposure to the technicalities of renormalization. The primary focus in this course lies on the treatment of UV divergences in conventional quantum field theories. Only the last lecture discusses renormalization in the context of effective field theories. We do not explore the structure of IR divergences in this course, since they are of a different origin. Let me just mention for completeness that effective theories have provided powerful new insights into the structure of IR divergences as well, see e.g.\ \cite{Becher:2009cu,Gardi:2009qi,Becher:2009qa,Becher:2009kw}.

\acknowledgements

I would like to thank my colleagues Sacha Davidson, Paolo Gambino and Mikko Laine for letting me deliver this lecture course despite of being one of the school organizers. Special thanks to Sacha for making sure that the students and lecturers were taken good care of at all times! I am grateful to the students for attending the lectures, asking lots of good questions and solving homework problems despite the busy schedule of the school. They have made delivering this course a true pleasure. During my stay at Les Houches I have enjoyed many interactions with my fellow lecturers, in particular with Thomas Becher, Aneesh Manohar and Toni Pich. I would also like to thank my students Stefan Alte, Javier Castellano Ruiz and Bianka Mecaj for careful proof-reading of these lecture notes and suggestions for improvements. 

This work has been supported by the Cluster of Excellence {\em Precision Physics, Fundamental Interactions and Structure of Matter\/} (PRISMA -- EXC 1098) at Johannes Gutenberg University Mainz. 

\tableofcontents

\maintext

\chapter{Renormalization in QED}
\label{sec:QED}

Loop diagrams in quantum field theories are plagued by ultraviolet (UV) divergences. The procedure of {\em renormalization\/} is a systematic way of removing these divergences by means of a {\em finite number\/} of redefinitions of the parameters of the theory. We will review this formalism first with the example of Quantum Electrodynamics (QED), the theory describing the interaction of electrically charged particles with light. For simplicity, we focus on the simplest version of the theory containing a single charged fermion, i.e.\ electrons and positrons. 

\section{UV divergences and renormalized perturbation theory}
\label{sec:UVdivs}

The Lagrangian of Quantum Electrodynamics (QED) reads (omitting gauge-fixing terms for simplicity)
\begin{equation}\label{LQED}
\begin{aligned}
   {\mathcal L}_{\rm QED} 
   &= \bar\psi_0\,(i\rlap{\,/}{D}-m_0)\,\psi_0 - \frac14\,F_{\mu\nu,0}\,F_0^{\mu\nu} \\
   &= \bar\psi_0\,(i\rlap{/}{\partial}-m_0)\,\psi_0 - \frac14\,F_{\mu\nu,0}\,F_0^{\mu\nu}
    - e_0\,\bar\psi_0\gamma_\mu\psi_0\,A_0^\mu \,,
\end{aligned}
\end{equation}
where $F_0^{\mu\nu}=\partial^\mu A_0^\nu-\partial^\nu A_0^\mu$ is the field-strength tensor. The Dirac field $\psi_0$ describes the electron and its anti-particle, the positron, and the vector field $A_0^\mu$ describes the photon. The parameters $m_0$ and $e_0$ account for the electron mass and its electric charge. We use a subscript ``0'' to distinguish the ``bare'' quantities appearing in the Lagrangian from the corresponding ``physical'' parameters -- i.e., the observable mass and electric charge of the electron -- and fields with proper (canonical) normalization. Renormalization theory yields the relations between the bare parameters and fields and the renormalized ones. 

By means of the Lehmann-Symanzik-Zimmermann (LSZ) reduction formula \cite{Lehmann:1954rq}, scattering amplitudes in quantum field theories are connected to fully connected, amputated Feynman diagrams. Moreover, we can restrict the following discussion to one-particle irreducible (1PI) graphs. One-particle reducible diagrams are simply products of 1PI graphs. A useful concept to classify the UV divergences of such diagrams is the so-called {\em superficial degree of divergence} $D$. For an arbitrary QED Feynman graph, the dependence on internal (unrestricted) momenta arises from the loop integrals and propagators:
\begin{equation}
   \begin{gathered}
   \includegraphics[scale=0.45]{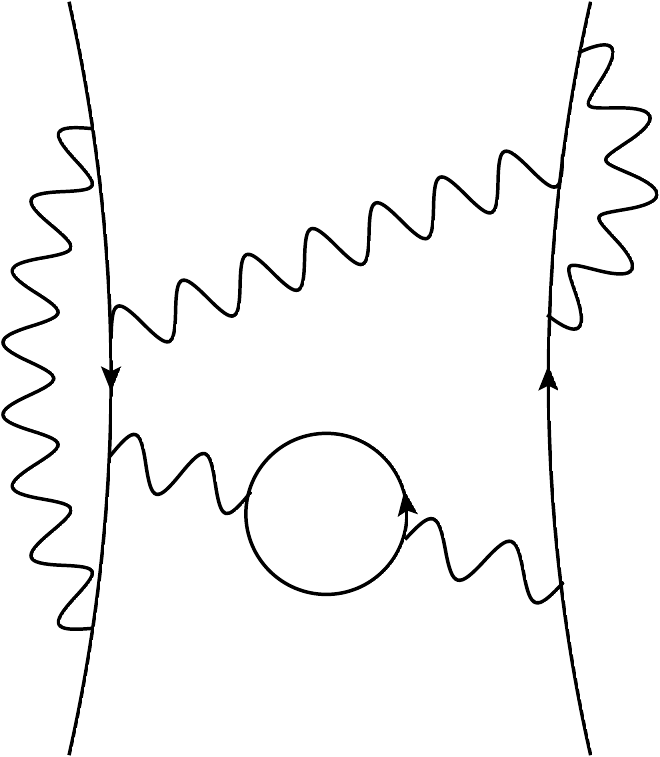}
   \end{gathered} 
   \sim \int\frac{d^4 k_1\dots d^4 k_L}{(\rlap/k_i-m+i0)\dots(k_j^2+i0)\dots}
\end{equation}
The quantity $D$ is defined as the sum of the powers of loop momenta in the numerator minus those in the denominator. Hence
\begin{equation}\label{Ddef}
   D = 4L - P_e - 2P_\gamma \,,
\end{equation}
where $L$ is the number of loops, and $P_e$ and $P_\gamma$ are the numbers of electron and photon propagators. One naively expects that diagrams with $D>0$ are {\em power divergent\/} ($\propto\Lambda_{\rm UV}^D$, where we denote by $\Lambda_{\rm UV}$ a generic UV cutoff regularizing the integral in the region of large loop momenta), diagrams with $D=0$ are {\em logarithmically divergent\/} ($\propto\ln\Lambda_{\rm UV}$), and diagrams with $D<0$ have no UV divergences. We will see below that in many cases the actual degree of divergence is less than $D$, as a consequence of gauge invariance or due to some symmetries. However, as long as we consider fully connected, amputated Feynman diagrams, the actual degree of divergence is never larger than $D$.

The beautiful combinatoric identity (problem~1.1)
\begin{equation}\label{beauty}
   L = I - V + 1
\end{equation}
relates the number of loops $L$ of any Feynman graph to the number of internal lines $I$ and the number of vertices $V$. For QED, this identity reads 
\begin{equation}
   L = P_e + P_\gamma - V + 1 \,.
\end{equation}
The only vertex of QED connects two fermion lines to a photon line, and hence we can express
\begin{equation}\label{vertex}
   V = 2P_\gamma + N_\gamma = \frac12\left( 2P_e + N_e \right) ,
\end{equation}
where $N_\gamma$ and $N_e$ denote the number of external photon and fermion lines, respectively. This equation follows since each propagator connects to two vertices, whereas each external line connects to a single vertex. Combining relations (\ref{Ddef}), (\ref{beauty}) and (\ref{vertex}), we obtain
\begin{equation}\label{DQED}
\begin{aligned}
   D &= 4 \left( P_e + P_\gamma - V + 1 \right) - P_e - 2P_\gamma \\
   &= 4 - 4V + 3P_e + 2P_\gamma \\[-0.5mm]
   &= 4 - 4V + 3 \left( V - \frac12\,N_3 \right) + \left( V - N_\gamma \right) \\[-2mm]
   &= 4 - \frac32\,N_e - N_\gamma \,.
\end{aligned}
\end{equation}
This relation is remarkable, since it relates the superficial degree of divergence of a graph to the number of external lines, irrespective of the internal complexity (the number of loops and vertices) of the graph. It follows that only a small number of $n$-point functions (sets of fully connected, amputated diagrams with $n$ external legs) have $D\ge 0$. It is instructive to look at them one by one (in each case, the blob represents infinite sets of graphs):

\[
\begin{array}{lr}
\parbox{3cm}{\hspace{8mm}\includegraphics[scale=0.36]{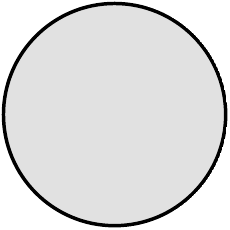}}
 & \parbox{10cm}{\small These so-called ``vacuum diagrams'' have $D=4$ and are badly divergent, but they give no contribution to $S$-matrix elements. As long as we ignore gravity, they merely produce an unobservable shift of the vacuum energy. (When gravity is taken into account, these graphs give rise to the infamous {\em cosmological constant problem}).} \\[12mm]
\parbox{3cm}{\hspace{8mm}\includegraphics[scale=0.36]{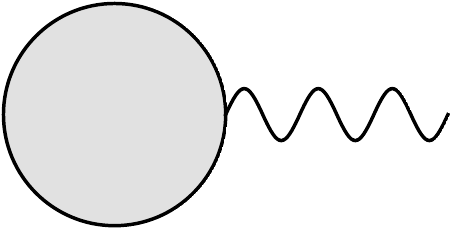}}
 & \parbox{10cm}{\small The one-photon amplitude has $D=3$, but it vanishes by Lorentz invariance. To see this, note that the amplitude (with the external polarization vector removed) has a Lorentz index $\mu$, but there is no 4-vector which could carry this index.} \\[10mm]
\parbox{3cm}{\hspace{0mm}\includegraphics[scale=0.36]{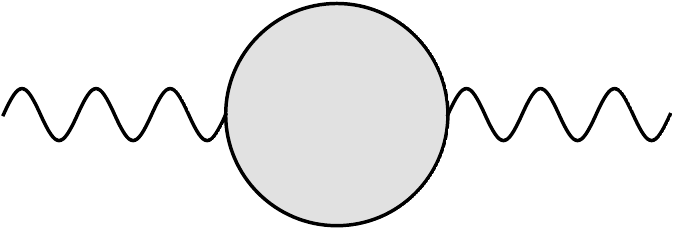}}
 & \parbox{10cm}{\small This so-called {\em photon vacuum polarization\/} amplitude has superficial degree of divergence $D=2$, and hence naively it is quadratically divergent. However, QED is a gauge theory, and gauge invariance requires that the vacuum polarization function has Lorentz structure $\pi^{\mu\nu}(k)=(k^2 g^{\mu\nu}-k^\mu k^\nu)\,\pi(k^2)$, see Section~\ref{subsec:2.2.2}. This means that two powers of loop momenta are replaced by external momenta, and hence the true degree of divergence is $D-2=0$, corresponding to a logarithmic UV divergence.} \\[14mm]
\parbox{3cm}{\hspace{0mm}\includegraphics[scale=0.36]{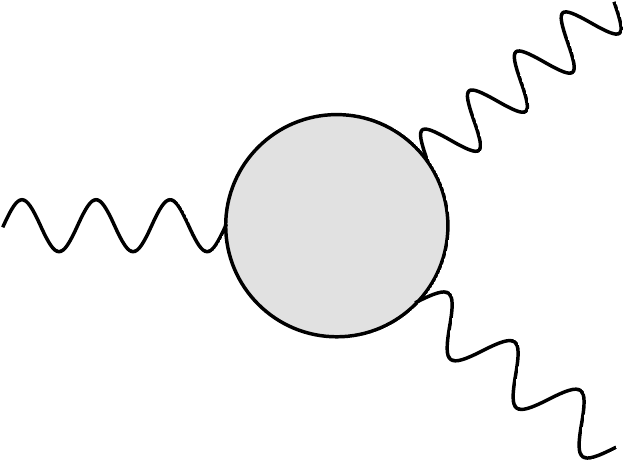}}
 & \parbox{10cm}{\small The three-photon amplitude has $D=1$, and hence naively it is linearly divergent. In QED, this amplitude vanishes as a result of invariance under $C$ parity (Furry's theorem). The same is true for all $(2n+1)$-photon amplitudes.} \\[9mm]
\parbox{3cm}{\hspace{1.8mm}\includegraphics[scale=0.36]{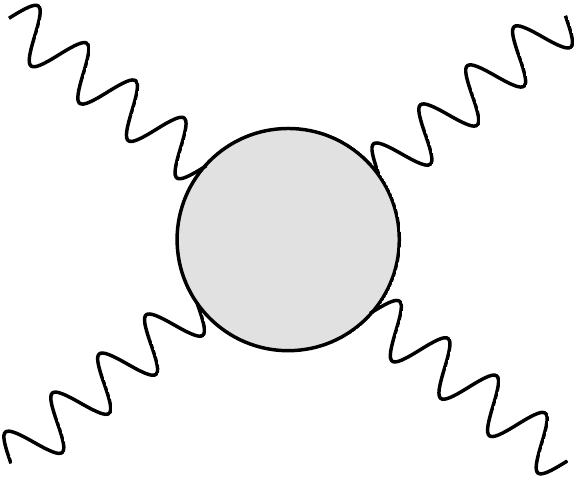}}
 & \parbox{10cm}{\small The four-photon amplitude has $D=0$, and hence naively it is logarithmically divergent. Due to gauge invariance, however, the amplitude involves four powers of external momenta (problem~1.2). Consequently, the true degree of divergence is $D-4=-4$, and so the amplitude is finite.} \\[12mm]
\parbox{3cm}{\includegraphics[scale=0.36]{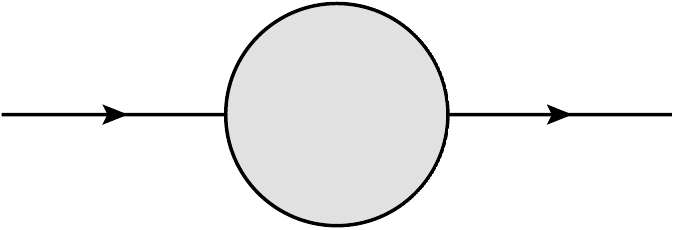}}
 & \parbox{10cm}{\small This so-called ``electron self energy'' has $D=1$, and hence naively it is linearly UV divergent. Chiral symmetry, i.e.\ the fact that in the limit $m_0=0$ left-handed and right-handed spinors transform under different irreducible representations of the Lorentz group, implies that the true degree of divergence is $D-1=0$, corresponding to a logarithmic UV divergence.} \\[7mm]
\parbox{3cm}{\hspace{4mm}\includegraphics[scale=0.36]{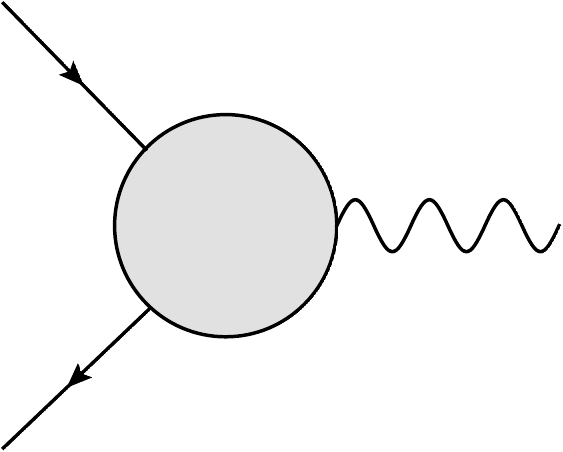}}
 & \parbox{10cm}{\small The electromagnetic vertex function has $D=0$ and is logarithmically UV divergent.} \\[5mm]
\end{array}
\]
\noindent
All other $n$-point functions in QED are UV finite.

Note that due to symmetries (Lorentz invariance, $C$ parity, gauge invariance, and the chiral symmetry of massless QED) the true degree of divergence is often less than the superficial degree of divergence $D$. In fact, the only divergent $n$-point functions are the two-point functions for the photon and the electron and the electromagnetic vertex function, which captures the quantum corrections to the only vertex of QED. This is a remarkable fact, which allows us to remove these divergences by multiplicative redefinitions of the electron and photon fields, the electron mass and the electric charge. We define the so-called ``renormalized'' fields (without subscript ``0'') by
\begin{equation}\label{WFR}
   \psi_0 = Z_2^{1/2}\,\psi \,, \qquad
   A_0^\mu = Z_3^{1/2}\,A^\mu \,.
\end{equation}
The renormalized fields will be chosen such that their two-point functions (the renormalized propagators) have unit residue (or, depending on the renormalization scheme, at least a finite residue) at $p^2=m^2$, where $m$ is the physical electron mass. The notation $Z_2$ and $Z_3$ for the renormalization factors is historical; it would probably make more sense to call them $Z_\psi$ and $Z_A$. When the Lagrangian (\ref{LQED}) is rewritten in terms of renormalized fields, one obtains  
\begin{equation}
   {\mathcal L}_{\rm QED} = Z_2\,\bar\psi\,(i\rlap{/}{\partial}-m_0)\,\psi
    - \frac{Z_3}{4}\,F_{\mu\nu}\,F^{\mu\nu}
    - Z_2\,Z_3^{1/2}\,e_0\,\bar\psi\gamma_\mu\psi\,A^\mu \,.
\end{equation}
In the next step, one relates the bare mass and electric charge to the corresponding physical quantities. Let us write the corresponding relations in the form
\begin{equation}\label{meren}
   Z_2\,m_0 = Z_m\,m \,, \qquad
   Z_2\,Z_3^{1/2}\,e_0 = \mu^{\frac{4-d}{2}} Z_1\,e \,.
\end{equation}
The scale $\mu$ enters in the dimensional regularization scheme \cite{Bollini:1972ui,tHooft:1972tcz}, in which the dimensionality of spacetime is analytically continued from 4 to $d<4$ (see Section~\ref{sec:2.2} below). It ensures that the renormalized charge $e$ is a dimensionless parameter. Expressed in terms of renormalized fields and parameters, the QED Lagrangian can now be written as
\begin{equation}\label{Lren}
\begin{aligned}
   {\mathcal L}_{\rm QED} 
   &= Z_2\,\bar\psi\,i\rlap{/}{\partial}\,\psi - Z_m\,m\,\bar\psi\psi
    - \frac{Z_3}{4}\,F_{\mu\nu}\,F^{\mu\nu} - \mu^{\frac{4-d}{2}} Z_1\,e\,\bar\psi\gamma_\mu\psi\,A^\mu \\
   &\equiv \bar\psi\,(i\rlap{/}{\partial} - m)\,\psi - \frac14\,F_{\mu\nu}\,F^{\mu\nu}
    - \mu^{\frac{4-d}{2}} e\,\bar\psi\gamma_\mu\psi\,A^\mu \\
   &\quad\mbox{}+ \bar\psi\,(\delta_2\,i\rlap{/}{\partial} - \delta m)\,\psi
    - \frac{\delta_3}{4}\,F_{\mu\nu}\,F^{\mu\nu}
    - \mu^{\frac{4-d}{2}} \delta_1\,e\,\bar\psi\gamma_\mu\psi\,A^\mu \,,
\end{aligned}
\end{equation}
where we have defined
\begin{equation}
\begin{aligned}
   \delta_2 &= Z_2 - 1 \,, \quad &\delta_3 &= Z_3 - 1 \,, \\
   \delta_1 &= Z_1 - 1 \,, \quad &\delta_m &= (Z_m-1)\,m \,.
\end{aligned}
\end{equation}
By construction, scattering amplitudes calculated from this Lagrangian, which are expressed in terms of the physical electron mass $m$ and electric charge $e$, are free of UV divergences. The first line in (\ref{Lren}) has the same structure as the original QED Lagrangian (apart from the factor $\mu^{\frac{4-d}{2}}$ in the electromagnetic vertex) and hence gives rise to the usual QED Feynman rules. If that was the entire story, we would still encounter UV-divergent results when computing Feynman graphs. However, the so-called ``counterterms'' in the second line give rise to additional Feynman rules, which have the effect of cancelling these UV divergences. The Feynman rules for these counterterms are as follows:
\begin{center}
\includegraphics[scale=0.4]{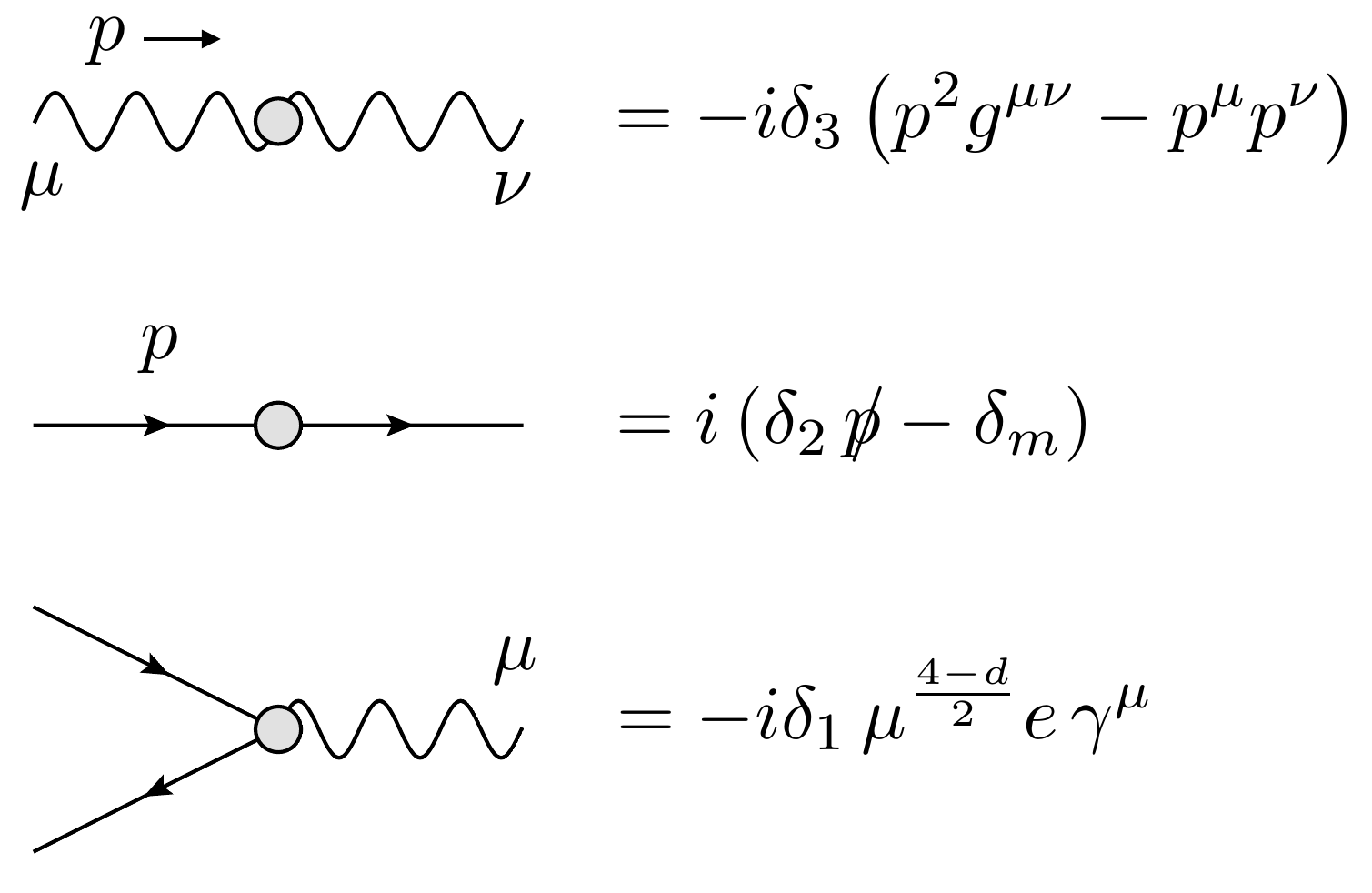}
\end{center}

The Lagrangian (\ref{Lren}) is the starting point for calculations in ``renormalized perturbation theory''\!, which gives rise to finite scattering amplitudes. The counterterms start at ${\mathcal O}(e^2)$ in perturbation theory and have a perturbative expansion in powers of the renormalized coupling $\alpha=e^2/(4\pi)$. Care must be taken to combine Feynman diagram with elementary vertices and counterterms at the same order in perturbation theory. When this is done consistently, the counterterms remove the UV divergences of Feynman graphs order by order in perturbation theory in $\alpha$. The proof of this statement is known as the Bogoliubov-Parasiuk-Hepp-Zimmermann (BPHZ) theorem \cite{Bogoliubov:1957gp,Hepp:1966eg,Zimmermann:1969jj}. It states that all divergences of quantum field theories can be removed by constructing counterterms for the superficially divergent 1PI Feynman graphs. For practical purposes, it is useful to note that renormalization works not only for entire $n$-point functions, but also for individual Feynman diagrams. Here are two examples:\\
\begin{center}
\includegraphics[scale=0.41]{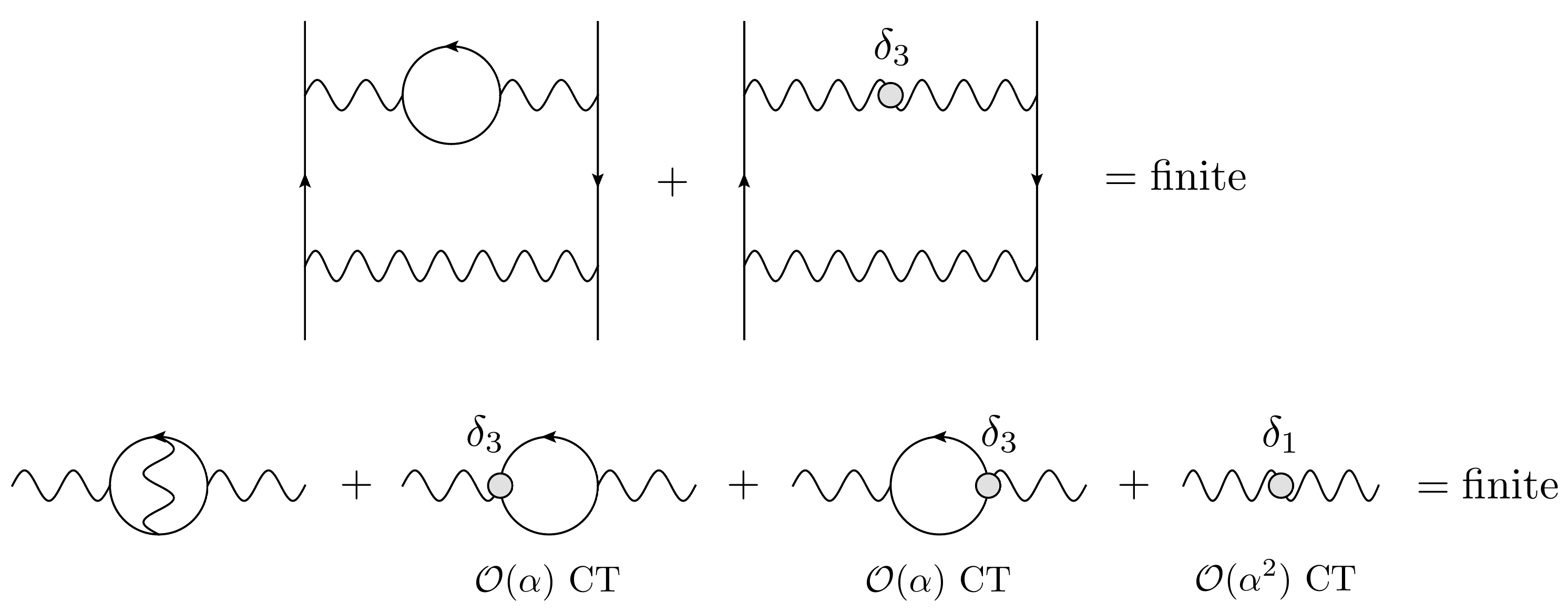}
\end{center}

\section{Calculation of the renormalization factors}
\label{sec:2.2}

We now understand that UV divergences only appear in intermediate steps of calculations in quantum field theories. When the counterterms are added to the bare Feynman graphs, these divergences cancel in all predictions for physical observables (e.g.\ scattering amplitudes). Nevertheless, in order to deal with the UV divergences arising in individual graphs, we must first introduce a regularization scheme. Ideally, the regularization should respect all symmetries of the theory as well as its fundamental properties, such as Lorentz invariance, gauge invariance, chiral symmetry (for $m_0=0$) and the analytic structure of scattering amplitudes. Also, the regulator should preserve the freedom to redefine the integration variables (the loop momenta). The at first sight most intuitive regularization scheme, in which we simply cut off loop integrals by means of a hard UV cutoff (such that $k_E^2<\Lambda_{\rm UV}^2$ after Wick rotation to Euclidean momenta), violates several of these requirements. In fact, the only known regularization scheme which preserves all of them is {\em dimensional regularization} \cite{Bollini:1972ui,tHooft:1972tcz}.\footnote{The Pauli--Villars scheme \cite{Pauli:1949zm} discussed in most textbooks on quantum field theory changes the analytic structure of scattering amplitudes and becomes cumbersome beyond one-loop order.} 
We have seen in the previous section that the UV divergences of QED $n$-point functions are at most of logarithmic strength. If we restrict the integrals over the loop momenta to less than~4 spacetime dimensions, then these logarithmically divergent integrals become finite. The ingenious idea of dimensional regularization is to take this observation serious. To this end, one replaces the four-dimensional loop integrals by $d$-dimensional ones:
\begin{equation}
   \int\frac{d^4k}{(2\pi)^4} \to \int\frac{d^dk}{(2\pi)^d} \quad \mbox{with} \quad d<4 \,.
\end{equation}
This lowers the degree of divergence of an $L$-loop diagram by $(d-4)L$, thus rendering logarithmically divergent integrals UV finite. We could now choose $d=3$ or some smaller integer value, but this would bring us to a lower-dimensional quantum field theory with very different properties than real-world QED. Instead, in dimensional regularization one considers an analytic continuation of spacetime to $d=4-2\epsilon$ dimensions, where $\epsilon>0$ is an infinitesimal parameter. In that way, the regularized theory we consider lives infinitesimally close to the original one.

If you have never been treated to a detailed exposition of dimensional regularization you will feel uncomfortable at this point. You are not alone in having problems imagining a $(4-2\epsilon)$-dimensional spacetime. The point is that using techniques we will briefly review below, loop integrals can be expressed in terms of analytic functions of the spacetime dimension $d$ with poles at integer values, reflecting singularities of the integral in $d$ (integer) dimensions. These functions can be {analytically continued\/} to the entire complex $d$-plane (which is more than we will need), in particular they can be continued to all real values of $d$. Since we need the dimensional regulator only in intermediate steps of the calculation, it is perfectly fine to work in the immediate vicinity of $d=4$, even if we cannot imagine what this means geometrically. UV singularities in 4 spacetime dimensions will show up as $1/\epsilon^n$ pole terms. When the counterterms are added to the original Feynman diagrams, these pole terms cancel and we can take the limit $\epsilon\to 0$ in the final result.

When the Lagrangian (\ref{LQED}) is continued to $d=4-2\epsilon$ spacetime dimensions, the canonical dimensions of the fields and parameters change. Using that the action $\int d^dx\,{\mathcal L}$ is dimensionless (as always in quantum field theory, we work in units where $\hbar=c=1$), it is straightforward to derive that (the brackets $[\dots]$ denote the mass dimension of a given quantity)  
\begin{equation}
   [\psi_0] = \frac{d-1}{2} = \frac32 - \epsilon \,, \quad
   [A_0^\mu] = \frac{d-2}{2} = 1 - \epsilon \,, \quad
   [m_0] = 1 \,, \quad
   [e_0] = \frac{4-d}{2} = \epsilon \,.
\end{equation}
If we wish to describe the strength of the electromagnetic interaction by means of a dimensionless coupling, we need to extract from the bare coupling $e_0$ a factor $\mu^\epsilon$ with some auxiliary mass scale $\mu$, as shown in (\ref{meren}), such that
\begin{equation}\label{e0rela}
   e_0\equiv \mu^\epsilon\,\tilde e_0(\mu) 
   = \mu^\epsilon\,Z_e(\mu)\,e(\mu) \,, \qquad Z_e = Z_1\,Z_2^{-1} Z_3^{-1/2} \,,
\end{equation}
where $\tilde e_0(\mu)$ is the dimensionless bare coupling and $e(\mu)$ is the renormalized coupling as defined in (\ref{meren}). Of course, physical quantities should not depend on the auxiliary scale $\mu$, which we have introduced for convenience only. As will be discussed later in Section~\ref{sec:2.3}, this fact gives rise to partial differential equations called {\em renormalization-group equations\/} (RGEs). 

Let me briefly mention a technical complication which will not be of much relevance to these lectures. Since the Clifford algebra $\{\gamma_\mu,\gamma_\nu\}=2g_{\mu\nu}$ involves the spacetime metric of Minkowski space, it needs to be generalized to $d$ dimensions when the dimensional regularization scheme is employed. It is not difficult to prove the following useful relations (problem~1.3):
\begin{equation}\label{Clifford}
   \gamma^\mu\gamma_\mu = d \,, \qquad
   \gamma^\mu\gamma_\alpha\gamma_\mu = (2-d)\,\gamma_\alpha \,, \qquad
   \gamma^\mu\gamma_\alpha\gamma_\beta\gamma_\mu = 4g_{\alpha\beta}
    + (d-4)\,\gamma_\alpha\gamma_\beta \,.
\end{equation}
In chiral gauge theories such as the Standard Model it is also necessary to generalize $\gamma_5$ to $d\ne 4$ dimensions. This is a problem full of subtleties, which will not be discussed here (see e.g.\ \cite{tHooft:1972tcz,Larin:1993tq} for more details). 

The evaluation of one-loop integrals (with loop momentum $k$) in dimensional regularization is a straightforward matter once one has learned a couple of basic techniques, which are taught in any textbook on quantum field theory. Let me briefly remind you of them:
\begin{enumerate}
\item
Combine the denominators of Feynman amplitudes, which contain products of propagators, using Feynman parameters. The general relation reads
\begin{equation}
   \frac{1}{A_1 A_2\cdots A_n}
   = \Gamma(n) \int_0^1\!dx_1\,\dots\int_0^1\!dx_n\,\delta\Big(1-\sum_{i=1}^n\,x_i\Big)\,
    \frac{1}{\left(x_1 A_1+\cdots+x_n A_n\right)^n} \,.
\end{equation}
Taking derivatives with respect to the $A_i$ allows one to derive analogous relations where the factors $A_i$ are raised to integer powers.
\item
Introduce a shifted loop momentum 
\begin{equation}
   \ell^\mu = k^\mu + \sum_i c_i(x_1,\dots,x_m)\,p_i^\mu \,,
\end{equation}
where $\{p_i^\mu\}$ are the external momenta of the diagram and the coefficients $c_i$ are linear functions of Feynman parameters, such that the integral takes on the standard form
\begin{equation}
   \int\frac{d^d\ell}{(2\pi)^d}\,\frac{1}{\left(\ell^2-\Delta+i0\right)^n}
   \left( N_0 + N_1\,\ell_\mu + N_2\,\ell_\mu\ell_\nu + \dots \right) .
\end{equation}
Note the absence of a linear term in $\ell$ in the denominator. The quantities $\Delta$ and $N_i$ depend on the Feynman parameters $\{x_i\}$ and the external momenta $\{p_i^\mu\}$.
\item
Use Lorentz invariance to replace
\begin{equation}
   \ell_\mu \to 0 \,, \qquad
   \ell_\mu\ell_\nu \to \frac{g_{\mu\nu}}{d}\,\ell^2 \,, \qquad
   \mbox{etc.}
\end{equation}
under the integral.
\item
The remaining integrals are performed using the Wick rotation $\ell^0\to i\ell_E^0$ (and hence $\ell^2\to-\ell_E^2$) and using spherical coordinates in $d$-dimensional Euclidean space. The relevant master formula reads
\begin{equation}
   \int\frac{d^d\ell}{(2\pi)^d}\,\frac{\left(\ell^2\right)^\alpha}{\left(\ell^2-\Delta+i0\right)^\beta}
   = \frac{i\,(-1)^{\alpha-\beta}}{(4\pi)^{\frac{d}{2}}} \left( \Delta-i0 \right)^{\alpha-\beta+\frac{d}{2}}\,
    \frac{\Gamma(\alpha+\frac{d}{2})\,\Gamma(\beta-\alpha-\frac{d}{2})}{\Gamma(\beta)\,\Gamma(\frac{d}{2})} \,.
\end{equation}
\item
Perform the integrals over the Feynman parameters $\{x_i\}$ either in closed form (if possible) or after performing a Laurent expansion about $\epsilon=0$. At one-loop order, the relevant parameter integrals can all be expressed in terms of logarithms and dilogarithms \cite{tHooft:1978jhc}.
\end{enumerate}
Let us now look at the results obtained for the three UV-divergent $n$-point functions of QED in the dimensional regularization scheme.

\subsection{Electron self energy}
\label{sec:2.2.1}

Consider the full electron propagator in momentum space. There are infinitely many diagrams contributing to this object, which we can classify by specifying the number of places in each diagram where the diagram falls apart if we cut a single electron line. Hence, the full propagator can be written as a geometric series of graphs containing more and more insertions of the so-called electron {\em self energy}, i.e., the infinite set of 1PI graphs with two external fermion legs:\\ 
\begin{equation}\label{fullprop}
\begin{aligned}
   &\begin{gathered}
    \includegraphics[scale=0.45]{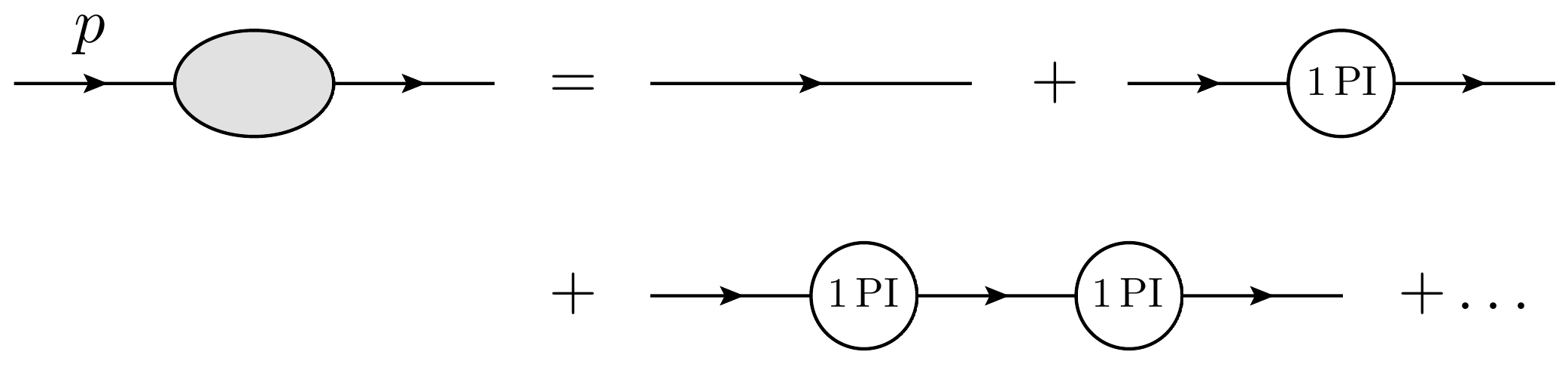}
    \end{gathered} \\
   &\quad = \frac{i}{\pslash-m_0+i0} 
    + \frac{i}{\pslash-m_0+i0} \left( -i \Sigma\right) \frac{i}{\pslash-m_0+i0} \\
   &\quad\quad\mbox{}+ \frac{i}{\pslash-m_0+i0} \left( -i \Sigma\right) \frac{i}{\pslash-m_0+i0} \left( -i \Sigma\right)
    \frac{i}{\pslash-m_0+i0} + \dots \\
   &\quad = \frac{i}{\pslash-m_0-\Sigma+i0} \,.
\end{aligned}
\end{equation}
The self energy $\Sigma\equiv\Sigma(\pslash,m_0,\alpha_0)$ can be expressed as a function of $\pslash$ and $p^2$, as well as of the bare parameters $m_0$ and $\alpha_0=\tilde e_0^2/(4\pi)$. Since $p^2=\pslash\pslash$, we do not need to list $p^2$ as an independent variable. The contributions to the self energy arising at one- and two-loop order in perturbation theory are:
\begin{center}
\includegraphics[scale=0.45]{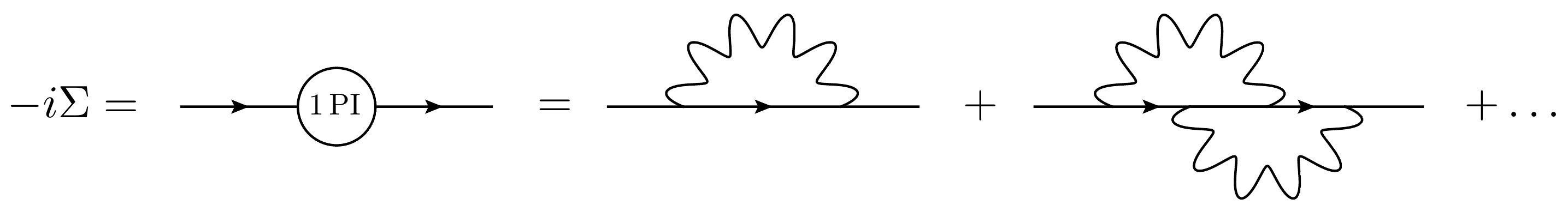}
\end{center}

The full propagator defined as the Fourier transform of the two-point function of two bare fermion fields has a pole  at the position of the physical electron mass $m$ with a residue equal to $Z_2$, the electron wave-function renormalization constant appearing in (\ref{WFR}): 
\begin{equation}\label{eq23}
   \begin{gathered}
    \includegraphics[scale=0.45]{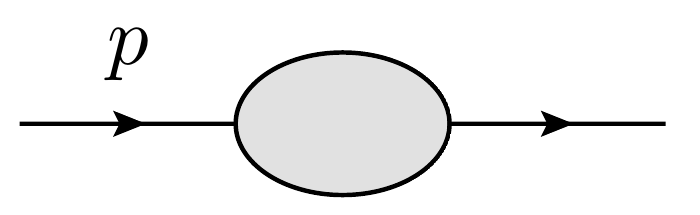}
    \end{gathered}
   ~ \stackrel{\mbox{\footnotesize $\rlap{\hspace{0.2mm}/}{p}\to m$}}{\,=\,}
    \frac{iZ_2}{\pslash-m+i0} + \mbox{less singular terms.}
\end{equation}
It follows that
\begin{equation}\label{basicrenorm}
\begin{aligned}
   m &= m_0 + \Sigma(\pslash=m,m_0,\alpha_0) \,, \\[1mm]
   Z_2^{-1} &= 1 - \frac{d\Sigma(\pslash=m,m_0,\alpha_0)}{d\pslash} 
    \bigg|_{\mbox{\footnotesize $\rlap{\hspace{0.2mm}/}{p}=m$}} \,.
\end{aligned}
\end{equation}
The action of the derivative operator $d/d\pslash$ on functions of $p^2$ is given by $df(p^2)/d\pslash=2\pslash\,f'(p^2)$. The first relation is an implicit equation for the renormalized mass $m$ in terms of the bare mass parameter $m_0$. At one-loop order, one finds (with Euler's constant $\gamma_E=0.5772\ldots$)
\begin{equation}\label{oneloop}
\begin{aligned}
   m &= m_0 \left[ 1 + \frac{3\alpha_0}{4\pi} \left( \frac{1}{\epsilon} - \gamma_E + \ln 4\pi
    + \ln\frac{\mu^2}{m_0^2} + \frac43 \right) + {\mathcal O}(\alpha_0^2) \right] , \\
   Z_2 &= 1 - \frac{\alpha_0}{4\pi} \left( \frac{1}{\epsilon} - \gamma_E + \ln 4\pi
    + \ln\frac{\mu^2}{m_0^2} - 2\ln\frac{m_0^2}{\lambda^2} + 4 \right) + {\mathcal O}(\alpha_0^2) \,.
\end{aligned}
\end{equation}
The derivative of the self energy evaluated at $\pslash=m$ is infrared (IR) divergent and gauge dependent. In this section we use the Feynman gauge ($\xi=1$) in the photon propagator
\begin{equation}
   D_F^{\mu\nu}(k) = \frac{-i}{k^2+i0} \left( g^{\mu\nu} - (1-\xi)\,\frac{k^\mu k^\nu}{k^2} \right)
\end{equation}
for simplicity. In the above expression for $Z_2$ we have regularized IR divergences by introducing a fictitious photon mass $\lambda$. IR divergences are not our main concern in these lectures, and hence we will not dwell on this issue further. 

The first relation in (\ref{oneloop}) appears to suggest that the physical mass $m$ is a quantity which diverges when one takes the limit $\epsilon\to 0$. However, we should instead write this equation as a relation for the bare mass parameter $m_0$ expressed in terms of the renormalized (and thus observable) mass $m$ and the renormalized coupling $\alpha=\alpha_0+{\mathcal O}(\alpha_0^2)$, such that
\begin{equation}\label{mrendef}
   m_0 = m \left[ 1 - \frac{3\alpha}{4\pi} \left( \frac{1}{\epsilon} - \gamma_E + \ln 4\pi
    + \ln\frac{\mu^2}{m^2} + \frac43 \right) + {\mathcal O}(\alpha^2) \right] .
\end{equation}
Likewise, we can rewrite the result for the wave-function renormalization constant of the electron in the from
\begin{equation}
   Z_2 = 1 - \frac{\alpha}{4\pi} \left( \frac{1}{\epsilon} - \gamma_E + \ln 4\pi
    + \ln\frac{\mu^2}{m^2} - 2\ln\frac{m^2}{\lambda^2} + 4 \right) + {\mathcal O}(\alpha^2) \,.
\end{equation}
The parameters $m$ and $\alpha$ on the right-hand side of these equations are measurable quantities. The equations tell us how the bare mass parameter $m_0$ and the normalization $Z_2$ of the bare fermion field diverge as the dimensional regulator $\epsilon=(4-d)/2$ is taken to zero. In Section~\ref{subsec:2.2.3} below, we will derive an analogous relation between the bare coupling constant $\alpha_0$ and the renormalized coupling $\alpha$.

The definitions (\ref{basicrenorm}) refer to the so-called on-shell renormalization scheme, in which $m=0.5109989461(31)$\,MeV is the physical mass of the electron \cite{Tanabashi:2018oca}, given by the pole position in the electron propagator, and in which $Z_2$ in the relation $\psi_0=Z_2^{1/2}\,\psi$ is defined such that the renormalized propagator defined as the Fourier transform of the two-point function $\langle\Omega|\,T\{\psi(x)\,\bar\psi(y)\}\,|\Omega\rangle$ has a unit residue at $\pslash=m$. In the next lecture we will introduce a different renormalization scheme, the so-called modified minimal subtraction ($\overline{\rm MS}$) scheme, in which the renormalized mass and residue will be defined in a different way.

For completeness, we also quote the renormalization factor of the electron mass as defined in the first relation in (\ref{meren}). We obtain
\begin{equation}
   Z_m = \frac{Z_2\,m_0}{m}
   = 1 - \frac{\alpha}{\pi} \left( \frac{1}{\epsilon} - \gamma_E + \ln 4\pi
    + \ln\frac{\mu^2}{m^2} - \frac12\,\ln\frac{m^2}{\lambda^2} + 2 \right) 
    + {\mathcal O}(\alpha^2) \,.
\end{equation}

\subsection{Photon vacuum polarization}
\label{subsec:2.2.2}

The self energy corrections for the gauge field are traditionally referred to as {\em vacuum polarization}. Consider the full photon propagator written as a series of contributions with more and more insertions of 1PI diagrams:
\begin{center}
\includegraphics[scale=0.45]{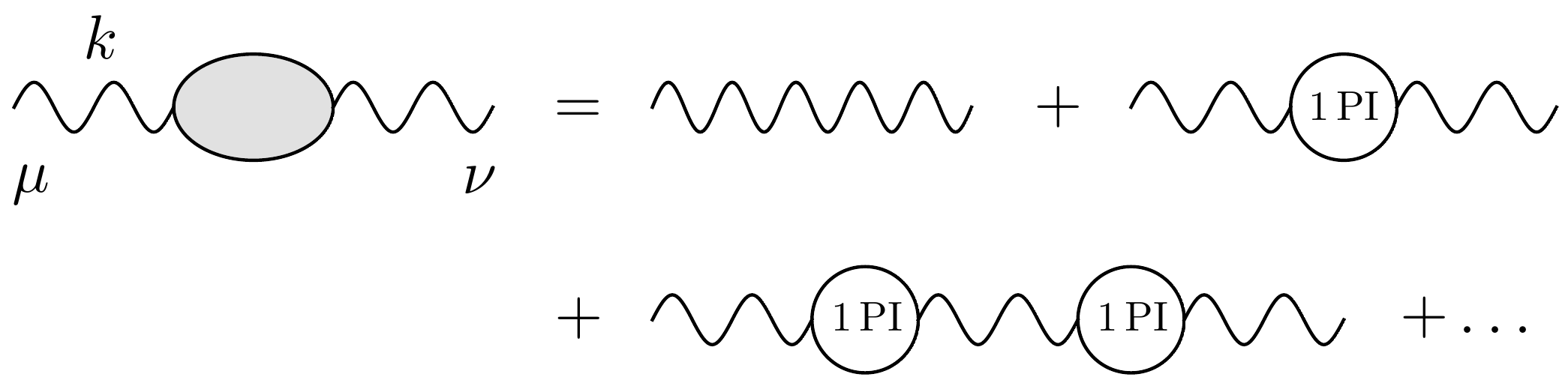}
\end{center}
Denote by $\pi^{\mu\nu}(k)$ the infinite set of 1PI propagator corrections. Up to two-loop order, the relevant diagrams are:
\begin{center}
\includegraphics[scale=0.38]{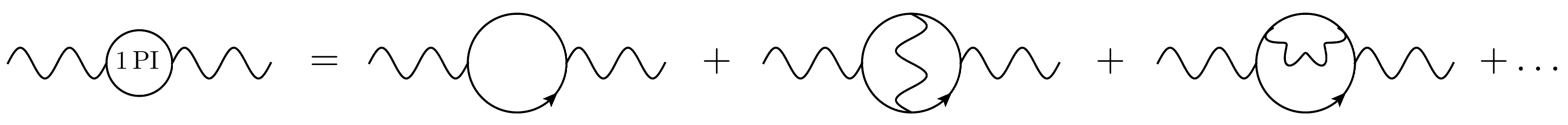}
\end{center}
Gauge invariance implies that $k_\mu\,\pi^{\mu\nu}(k)=k_\nu\,\pi^{\mu\nu}(k)=0$, and hence
\begin{equation}\label{pimunu}
   \pi^{\mu\nu}(k) = \left( g^{\mu\nu} k^2 - k^\mu k^\nu \right) \pi(k^2) \,.
\end{equation}
Performing the sum of the geometric series in an arbitrary covariant gauge, one obtains for the full photon propagator (problem~1.4)
\begin{equation}\label{Aprop}
   \begin{gathered}
    \includegraphics[scale=0.45]{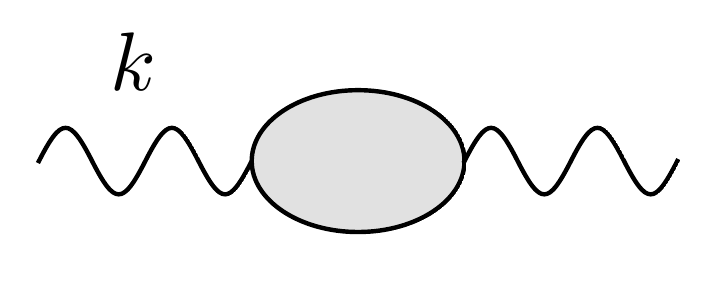}
    \end{gathered}
   ~ = \frac{-i}{k^2\left[ 1 - \pi(k^2) \right] + i0} 
    \left( g^{\mu\nu} - \frac{k^\mu k^\nu}{k^2+i0} \right) 
    - i\xi\,\frac{k^\mu k^\nu}{\left( k^2+i0 \right)^2} .
\end{equation}
Here $\xi$ is the gauge parameter ($\xi=1$ in Feynman gauge). Remarkably, the quantum corrections only affect the first term on the right-hand side, which contains the physical (transverse) polarization states. Also, as long as the function $\pi(k^2)$ is regular at the origin, these corrections do not shift the pole in the propagator. Indeed, the full propagator has a pole at $k^2=0$ with residue
\begin{equation}
   Z_3 = \frac{1}{1-\pi(0)} \,.
\end{equation}
From the relevant one-loop diagram, one obtains
\begin{equation}\label{Z3res}
\begin{aligned}
   Z_3 &= 1 - \frac{\alpha_0}{3\pi} \left( \frac{1}{\epsilon} - \gamma_E + \ln 4\pi
    + \ln\frac{\mu^2}{m_0^2} \right) + {\mathcal O}(\alpha_0^2) \\
   &= 1 - \frac{\alpha}{3\pi} \left( \frac{1}{\epsilon} - \gamma_E + \ln 4\pi
    + \ln\frac{\mu^2}{m^2} \right) + {\mathcal O}(\alpha^2) \,.
\end{aligned}
\end{equation}

\subsection{Charge renormalization}
\label{subsec:2.2.3}

Besides the electron and photon propagators, our analysis in Section~\ref{sec:UVdivs} had indicated that the electromagnetic vertex function coupling a photon to an electron--positron pair contains UV divergences, too. In momentum space, and using the Gordon identity, the vertex function can be written as
\begin{equation}
   -i\tilde e_0\,\Gamma^\mu(p,p') 
   = -i\tilde e_0 \left[ \gamma^\mu\,\Gamma_1(q^2) + \frac{i\sigma^{\mu\nu} q_\nu}{2m}\,\Gamma_2(q^2) \right] ,
\end{equation}
where $q^\mu=(p'-p)^\mu$ is the momentum transfer, and $\Gamma^\mu(p,p')$ includes the 1PI vertex-correction graphs, i.e.:
\begin{center}
\includegraphics[scale=0.45]{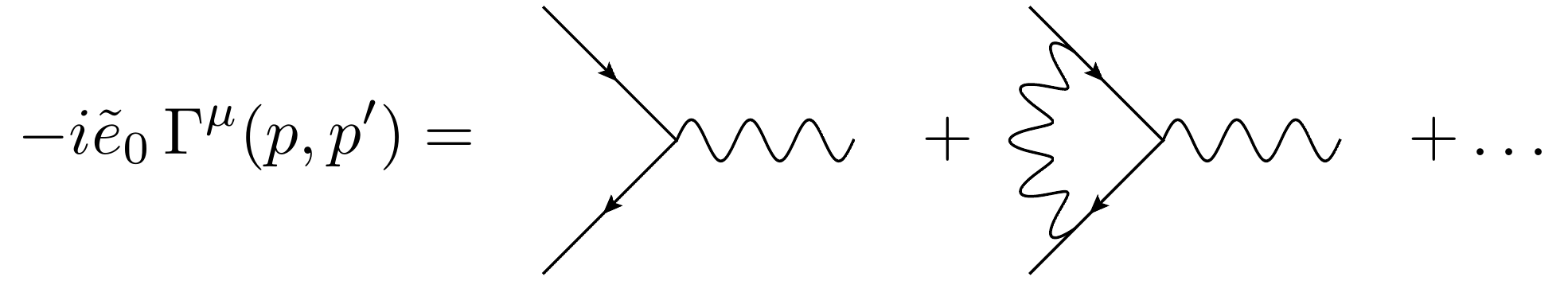}
\end{center}
Only the structure $\Gamma_1(q^2)$ is UV divergent. In the on-shell renormalization scheme, one defines
\begin{equation}
   Z_1 = [\Gamma_1(0)]^{-1} \,.
\end{equation}
The Ward--Takahashi identity of QED \cite{Ward:1950xp,Takahashi:1957xn} 
\begin{equation}\label{Ward}
   -iq_\mu\,\Gamma^\mu(p,p') = S^{-1}(p') - S^{-1}(p) \,,
\end{equation}
where $S(p)$ denotes the full electron propagator in (\ref{fullprop}), implies that $Z_1=Z_2$ to all orders of perturbation theory, where we have used (\ref{eq23}). From (\ref{meren}), we then obtain the following relation between the bare and the renormalized electric charges:
\begin{equation}
   e_0 = \mu^\epsilon\,Z_1\,Z_2^{-1} Z_3^{-1/2}\,e = \mu^\epsilon\,Z_3^{-1/2}\,e \,.
\end{equation}
For the coupling $\alpha_0=\tilde e_0^2/(4\pi)$, this relation implies
\begin{equation}\label{alpharendef}
   \alpha_0 = \alpha \left[ 1 + \frac{\alpha}{3\pi} \left( \frac{1}{\epsilon} - \gamma_E + \ln 4\pi
    + \ln\frac{\mu^2}{m^2} \right) + {\mathcal O}(\alpha^2) \right] \,.
\end{equation}
Here $\alpha=1/137.035999139(31)$ is the {\em fine-structure constant\/} \cite{Tanabashi:2018oca}, defined in terms of the photon coupling to the electron at very small momentum transfer ($q^2\to 0$). It is one of the most precisely known constants of nature. This concludes the calculation of the renormalization constants of QED.

\subsection{Counterterms}

Given the above results, it is straightforward to derive the one-loop expressions for the counterterms of QED in the dimensional regularization scheme. We find
\begin{equation}
\begin{aligned}
   \delta_1 = \delta_2 &= - \frac{\alpha}{4\pi} \left( \frac{1}{\epsilon} - \gamma_E + \ln 4\pi
    + \ln\frac{\mu^2}{m^2} - 2\ln\frac{m^2}{\lambda^2} + 4 \right) + {\mathcal O}(\alpha^2) \,, \\
   \delta_3 &= - \frac{\alpha}{3\pi} \left( \frac{1}{\epsilon} - \gamma_E + \ln 4\pi
    + \ln\frac{\mu^2}{m^2} \right) + {\mathcal O}(\alpha^2) \,, \\
   \delta_m &= - \frac{\alpha}{\pi} \left( \frac{1}{\epsilon} - \gamma_E + \ln 4\pi
    + \ln\frac{\mu^2}{m^2} - \frac12\,\ln\frac{m^2}{\lambda^2} + 2 \right) 
    + {\mathcal O}(\alpha^2) \,.
\end{aligned}
\end{equation}

\section{Scale dependence in the on-shell renormalization scheme}
\label{sec:2.3}

So far we have worked in the on-shell renormalization scheme, in which the renormalized mass and coupling constant are related to well measured physical constants (the electron mass and the fine-structure constant), and in which the renormalized fields are defined such that the renormalized propagators have poles with unit residues at the physical masses. For most calculations in QED (as well as for many calculations in the theory of electroweak interactions) this is the most convenient renormalization scheme.

You might be confused by the following subtlety related to the definition of the renormalized parameters in the on-shell scheme. Clearly, the bare parameters $m_0$ and $e_0$ in the QED Lagrangian are independent of the auxiliary scale $\mu$, which we have introduced in (\ref{meren}). At first sight, it appears that the renormalized mass and coupling defined in (\ref{mrendef}) and (\ref{alpharendef}) must be scale-dependent quantities. But I just told you that these parameters have been measured with high precision and thus they are definitely independent of $\mu$. The resolution of this puzzle rests on the fact that the relations between the bare and renormalized quantities are defined in the regularized theory in $d=4-2\epsilon$ spacetime dimensions. While in any renormalizable quantum field theory it is possible to take the limit $\epsilon\to 0$ at the end of a calculation of some observable, this limit must not be taken in relations such as (\ref{mrendef}) and (\ref{alpharendef}), since the bare parameters $m_0$ and $e_0$ would diverge in this limit. Using the fact that in the on-shell scheme both the bare and renormalized parameters are $\mu$ independent {\em after\/} we take $\epsilon\to 0$, it follows from (\ref{meren}) and (\ref{e0rela}) that
\begin{equation}\label{onshell}
\begin{aligned}
   \mu\,\frac{d}{d\mu}\,\Big[ \mu^\epsilon Z_e\,e(\mu) \Big] 
   &= \left( \mu\,\frac{d}{d\mu}\,Z_e \right) \mu^\epsilon\,e(\mu)
    + \mu^\epsilon Z_e \left[ \epsilon\,e(\mu) + \mu\,\frac{d}{d\mu}\,e(\mu) \right] = 0 \,, \\
   \mu\,\frac{d}{d\mu}\,\frac{Z_m}{Z_2}
   &= \left( \frac{\partial}{\partial\ln\mu} + \mu\,\frac{de(\mu)}{d\mu}\,\frac{\partial}{\partial e}
    \right) \frac{Z_m}{Z_2} = 0 \,,
\end{aligned}
\end{equation}
where $Z_e=Z_3^{-1/2}$. In the on-shell renormalization scheme (but not in other schemes!), the first relation is solved by
\begin{equation}
   \mu\,\frac{d}{d\mu}\,e(\mu) = - \epsilon\,e(\mu) \,, \qquad    
   \mu\,\frac{d}{d\mu}\,Z_e 
   = \left( \frac{\partial}{\partial\ln\mu}
    - \epsilon\,e\,\frac{\partial}{\partial e} \right) Z_e = 0 \,.
\end{equation}
In terms of the renormalized coupling $\alpha(\mu)$, this becomes
\begin{equation}\label{eq1.42}
   \mu\,\frac{d}{d\mu}\,\alpha(\mu) = - 2\epsilon\,\alpha(\mu) \,, \qquad    
   \mu\,\frac{d}{d\mu}\,Z_e 
   = \left( \frac{\partial}{\partial\ln\mu}
    - 2\epsilon\,\alpha\,\frac{\partial}{\partial\alpha} \right) Z_e = 0 \,.
\end{equation}
The first relation states that in the regularized theory in $d=4-2\epsilon$ dimensions the renormalized coupling is indeed scale dependent, but its scale dependence is simply such that $\alpha(\mu)\propto\mu^{-2\epsilon}$. Once we take the limit $\epsilon\to 0$ at the end of a calculation, the renormalized coupling in the on-shell scheme becomes a scale-independent constant, i.e.\ 
\begin{equation}
   \lim_{\epsilon\to 0}\,\alpha(\mu) = \alpha = \frac{1}{137.035999139(31)} \,.
\end{equation}
Using that $Z_e=Z_3^{-1/2}$ with $Z_3$ given in (\ref{Z3res}), it is straightforward to check that the second relation in (\ref{eq1.42}) is indeed satisfied at one-loop order. Finally, the second relation in (\ref{onshell}) translates into 
\begin{equation}
   \left( \frac{\partial}{\partial\ln\mu} - 2\epsilon\,\alpha\,\frac{\partial}{\partial\alpha}
    \right) \frac{Z_m}{Z_2} = 0 \,.
\end{equation}
It is again easy to see that this relation holds.

\section{Renormalization schemes}

While the on-shell scheme is particularly well motivated physically, it is not the only viable renormalization scheme. The only requirement we really need to ask for is that the counter\-terms remove the UV divergences of Feynman diagrams. The minimal way of doing this is to include {\em only\/} the $1/\epsilon^n$ pole terms in the counterterms (where in most cases $n=1$ at one-loop order) and leave all finite terms out. This is referred to as the minimal subtraction (MS) scheme \cite{tHooft:1973mfk,Weinberg:1951ss}. In fact, since as we have seen the $1/\epsilon$ poles always come along with an Euler constant and a logarithm of $4\pi$, it is more convenient to remove the poles in
\begin{equation}
   \frac{1}{\hat\epsilon}\equiv \frac{1}{\epsilon} - \gamma_E + \ln 4\pi \,.
\end{equation}
The corresponding scheme is called the modified minimal subtraction ($\overline{\rm MS}$) scheme \cite{Bardeen:1978yd}, and it is widely used in perturbative calculations in high-energy physics and in QCD in particular. Let us summarize the QED renormalization factors in the $\overline{\rm MS}$ scheme. We have
\begin{equation}\label{ZiMSbar}
\begin{aligned}
   Z_1^{\overline{\rm MS}} = Z_2^{\overline{\rm MS}} 
   &= 1 - \frac{\alpha}{4\pi\hat\epsilon} + {\mathcal O}(\alpha^2) \,, \\
   Z_3^{\overline{\rm MS}} &= 1 - \frac{\alpha}{3\pi\hat\epsilon} + {\mathcal O}(\alpha^2) \,, \\
   Z_m^{\overline{\rm MS}} &= 1 - \frac{\alpha}{\pi\hat\epsilon} + {\mathcal O}(\alpha^2) \,.
\end{aligned}
\end{equation}

The renormalized electron mass and charge in the $\overline{\rm MS}$ scheme are free of divergences, but these quantities are {\em no longer scale independent}. In fact, it is straightforward to relate these parameters to those defined in the on-shell scheme. We obtain
\begin{equation}\label{MSbarrun}
\begin{aligned}
   m_{\overline{\rm MS}}(\mu) &= m\,\frac{(Z_2/Z_m)^{\overline{\rm MS}}}{(Z_2/Z_m)^{\rm OS}}
    = m \left[ 1 - \frac{3\alpha}{4\pi} \left( \ln\frac{\mu^2}{m^2} + \frac43 \right) + {\mathcal O}(\alpha^2) \right] , \\
   \alpha_{\overline{\rm MS}}(\mu) &= \alpha\,\frac{Z_3^{\overline{\rm MS}}}{Z_3^{\rm OS}}
    = \alpha \left( 1 + \frac{\alpha}{3\pi}\,\ln\frac{\mu^2}{m^2} + {\mathcal O}(\alpha^2) \right) .
\end{aligned}
\end{equation}
While it may appear inconvenient at first sight to express the results of calculations in quantum field theory in terms of such scale-dependent (or ``running'') parameters, we will encounter situations where this is indeed very useful. As a rule of thumb, this is always the case when the characteristic energy or mass scale of a process is much larger than the electron mass. The running electron mass $m_{\overline{\rm MS}}(\mu)$ decreases with increasing $\mu$, while the running coupling $\alpha_{\overline{\rm MS}}(\mu)$ increases. Note that physical observables such as cross sections for scattering events or decay rates of unstable particles are always scale independent, i.e.\ the scale dependence of the running parameters is compensated by scale-dependent terms in the perturbative series for these quantities. This will be discussed in more detail in Section~\ref{sec:RGEs}.

\section{Homework problems}

\begin{enumerate}
\item[1.1]
Prove the combinatoric identity (\ref{beauty}).  
\item[1.2]
Gauge invariance requires that the four-photon amplitude $\pi^{\alpha\beta\gamma\delta}(k_1,k_2,k_3,k_4)$ (without external polarization vectors, and with incoming momenta satisfying $k_1+k_2+k_3+k_4=0$) vanishes when one of its Lorentz indices is contracted with the corresponding external momentum vector, e.g.\ $k_{1\alpha}\,\pi^{\alpha\beta\gamma\delta}(k_1,k_2,k_3,k_4)=0$. Use this fact as well as Bose symmetry to derive the most general form-factor decomposition of this amplitude and show that the amplitude is UV finite. 
\item[1.3]
Prove the relations (\ref{Clifford}) for the $d$-dimensional Dirac matrices using the Clifford algebra $\{\gamma_\mu,\gamma_\nu\}=2g_{\mu\nu}$.  
\item[1.4]
Derive relation (\ref{Aprop}) for the full photon propagator.
\end{enumerate}

\chapter{Renormalization in QCD}

The Lagrangian of Quantum Chromodynamics (QCD), the fundamental theory of the strong interactions, is structurally very similar to the QED Lagrangian in (\ref{LQED}). For the case of a single flavor of quarks, it reads (omitting gauge-fixing terms for simplicity)
\begin{equation}\label{LQCD}
   {\mathcal L}_{\rm QCD} 
   = \bar\psi_{q,0}\,(i\rlap{\,/}{D}-m_{q,0})\,\psi_{q,0} - \frac14\,G_{\mu\nu,0}^a\,G_0^{\mu\nu,a} 
    + \bar c_0^a\,(-\partial^\mu D_\mu^{ab})\,c_0^b \,.
\end{equation}
Here $\psi_q$ is the Dirac spinor for the quark field, $m_q$ denotes the quark mass, $G_{\mu\nu}^a$ is the field-strength tensor of the gluon fields, and $c^a$ are the Faddeev--Popov ghost fields. As before the subscript ``0'' is used to indicate ``bare'' quantities in the Lagrangian.

In the real world, QCD contains six different types of quark fields with different masses, referred to as ``flavors'', which are called up, down, strange, charm, bottom (or beauty), and top (or truth). Strictly speaking, a sum over quark flavors should thus be included in the above Lagrangian.\footnote{Likewise, in the real world there exist three types of charged leptons, called the electron, the muon and the tau lepton. In our discussion in Section~\ref{sec:QED} we have ignored the presence of the muon and the tau lepton, which have masses much heavier than the electron.}   

The main differences between QED and QCD are due to the fact that QCD is a {\em non-abelian\/} gauge theory based on the group $SU(N_c)$, where $N_c=3$ is called the number of colors. While in QED particles carry a single charge (the electric charge $\pm e$ or a fraction thereof), the quarks carry one of three colors $i=1,2,3$. In fact, quarks live in the fundamental representation of the gauge group, and the quark spinor field $\psi_q$ can be thought of as a 3-component vector in color space. The gluons, the counterparts of the photon in QED, live in the adjoint representation of the gauge group, which is $(N_c^2-1)$ dimensional. Hence there are eight gluon fields in QCD, labeled by an index $a=1,\dots,8$. When acting on quark fields, the covariant derivative reads
\begin{equation}
   iD_\mu = i\partial_\mu + g_s\,A_\mu^a\,t^a \,,
\end{equation}
where the eight $3\times 3$ matrices $t^a$ are called the Gell-Mann matrices. They are the generators of color rotations in the fundamental representation. (When $D_\mu$ acts on the ghost fields, the generators $t^a$ must instead be taken in the adjoint representation.) The strong coupling $g_s$ replaces the electromagnetic coupling $e$ in QED. The most important difference results from the form of the QCD field-strength tensor, which reads
\begin{equation}\label{Gmunu}
   G_{\mu\nu}^a = \partial_\mu A_\nu^a - \partial_\nu A_\mu^a + g_s\,f_{abc}\,A_\mu^b A_\nu^c \,.
\end{equation}
The quadratic term in the gauge potentials arises since the commutator of two color generators is non-vanishing, \begin{equation}
   [t_a,t_b] = i f_{abc}\,t_c \,,
\end{equation}
and it thus reflects the non-abelian nature of the gauge group. 

Let us briefly summarize the main differences between QED and QCD, all of which result from the differences between the abelian group $U(1)$ and the non-abelian group $SU(N_c)$:
\begin{enumerate}
\item
In QED there is a single elementary vertex connecting two electron lines to a photon line. A similar vertex coupling two quark lines to a gluon line also exists in QCD. However, because of the structure of (\ref{Gmunu}), there are in addition gluon self-interactions connecting three or four gluons at a single vertex. 
\item
The gauge-fixing procedure in non-abelian gauge theories gives rise to a non-trivial functional determinant, which is dealt with by introducing Faddeev--Popov ghost fields $c^a$. These are anti-commuting scalar fields transforming in the adjoint representation, i.e., fields with the wrong spin-statistics relation, which hence cannot appear as external states in scattering amplitudes. Internal ghost fields inside Feynman graphs such as 
\begin{center}
\includegraphics[scale=0.42]{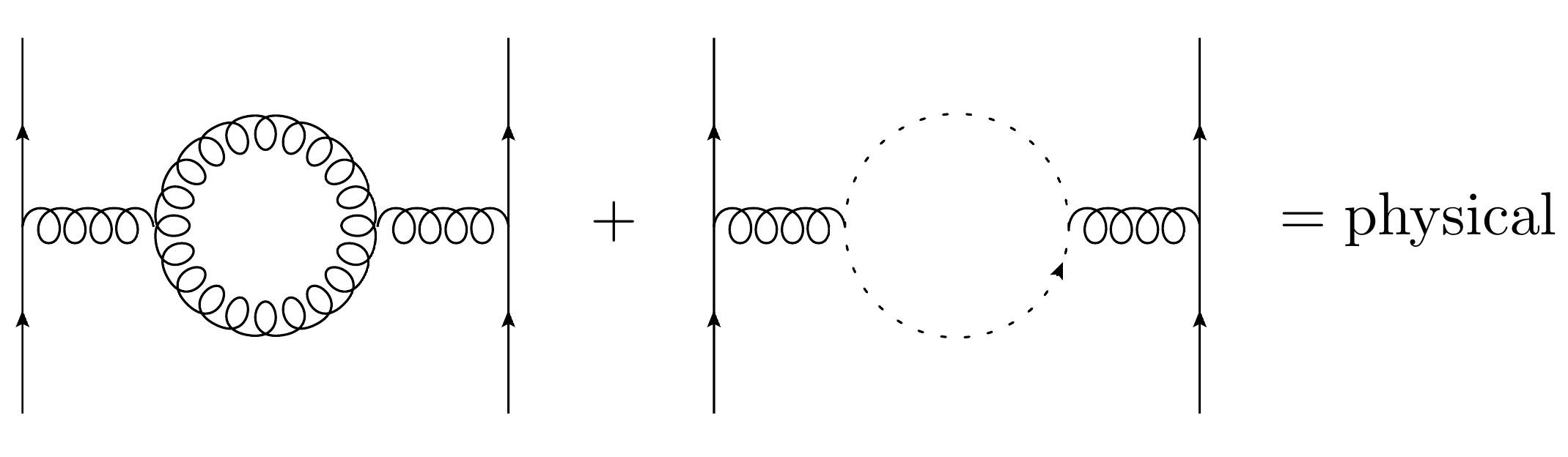}
\end{center}
are however needed to cancel the unphysical gluon polarizations in loops. The presence of the ghost fields gives rise to an additional elementary vertex connecting two ghost lines to a gluon line.
\item
Unlike in QED, in QCD calculations one encounters non-trivial group-theory factors, the most common ones being
\begin{equation}
   C_F = \frac{N_c^2-1}{2N_c} = \frac43 \,, \qquad
   C_A = N_c = 3 \,, \qquad
   T_F = \frac12 \,.
\end{equation}
Important relations involving these factors are (summed over repeated indices) 
\begin{equation}\label{colors}
   t_a\,t_a = C_F\,\bm{1} \,, \qquad
   f_{abc}\,f_{abd} = C_A\,\delta_{cd} \,, \qquad
   \mbox{Tr}(t_a\,t_b) = T_F\,\delta_{ab} \,.
\end{equation}
\item
The superficial degree of divergence of a 1PI Feynman diagram in QED has been given in (\ref{DQED}). For a 1PI graph in QCD, one can prove that (problem~2.1)
\begin{equation}\label{DQCD}
   D = 4 - \frac32\,N_q - N_g - \frac32\,N_c \,,
\end{equation}
where $N_q$, $N_g$ and $N_c$ are the number of external quark, gluon and ghost lines. Note that, while scattering amplitudes cannot contain external ghost particles, there do exist UV-divergent 1PI vertex functions involving external ghost fields.
\end{enumerate}
Perhaps the most important difference concerns the phenomenology of the two theories. While QED is a weakly coupled quantum field theory for all relevant energy scales,\footnote{QED would get strongly coupled near the Landau pole of the running coupling $\alpha(\mu)$ in (\ref{MSbarrun}), which however lies far above the Planck scale.} 
QCD exhibits strong-coupling behavior at low energies but weak-coupling behavior at high energies (``asymptotic freedom''). The strong coupling at low energies gives rise to the phenomenon of {\em color confinement}, which is the statement that in the low-energy world quarks and gluons are always locked up inside colorless bound states called hadrons. 

\section{Renormalization in QCD}

While the on-shell renormalization scheme is useful for many (but not all) calculations in QED, it is {\em not\/} a viable renormalization scheme for QCD calculations, for the following reasons: 
\begin{itemize}
\item
Quarks and gluons can never be on-shell because of confinement. In the real world free (isolated) quarks and gluons do not exist, and hence the corresponding two-point functions do not have poles at $p^2=m_q^2$ or $k^2=0$, respectively.
\item
The strong coupling $g_s$ and the associated parameter $\alpha_s=g_s^2/(4\pi)$ cannot be renormalized at $q^2=0$, since QCD is strongly coupled at low or vanishing momentum transfer and hence quarks and gluons are not the relevant degrees of freedom to describe the strong interactions in this regime.
\item
The masses of the three light quark flavors satisfy $m_q\ll\Lambda_{\rm QCD}$ (for $q=u,d,s$), where $\Lambda_{\rm QCD}$ is (roughly) the scale at which QCD becomes strongly coupled. It is therefore a good approximation for many purposes to set the light quark masses to zero. We will not consider heavy quarks with masses $m_Q\gg\Lambda_{\rm QCD}$ (for $Q=c,b,t$) in these lectures. The effects of heavy quarks are usually described using some kind of effective field theory. This is discussed in detail in the lecture courses by Thomas Mannel, Luca Silvestrini and Rainer Sommer elsewhere in this book. 
\end{itemize}
For all these reasons, one uses the $\overline{\rm MS}$ renormalization scheme for perturbation-theory calculations in QCD.

Let us briefly discuss the structure of UV-divergent vertex functions in QCD. In addition to the analogues of the divergent $n$-point functions in QED, the following amplitudes which arise only in QCD are UV divergent and require renormalization:\\[-6mm]

\[
\begin{array}{lr}
\parbox{3cm}{\hspace{8mm}\includegraphics[scale=0.36]{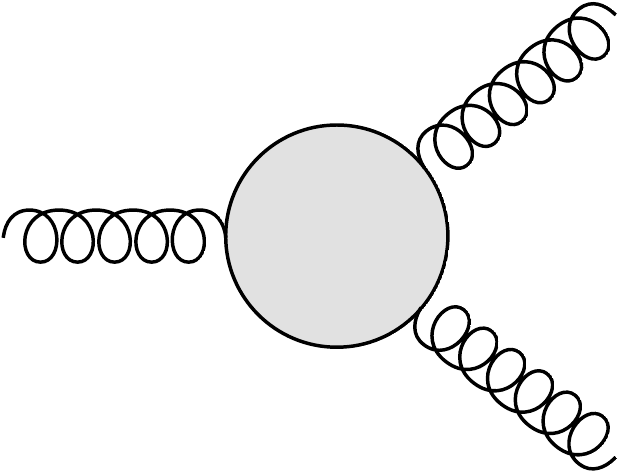}}
 & \parbox{10cm}{\small The three-gluon amplitude has $D=1$, and hence naively it is linearly divergent. Unlike in QED, in QCD this amplitude no longer vanishes (i.e.\ Furry's theorem does not apply in QCD), but it only contains logarithmic UV divergences due to gauge invariance.} \\[11mm]
\parbox{3cm}{\hspace{2mm}\includegraphics[scale=0.36]{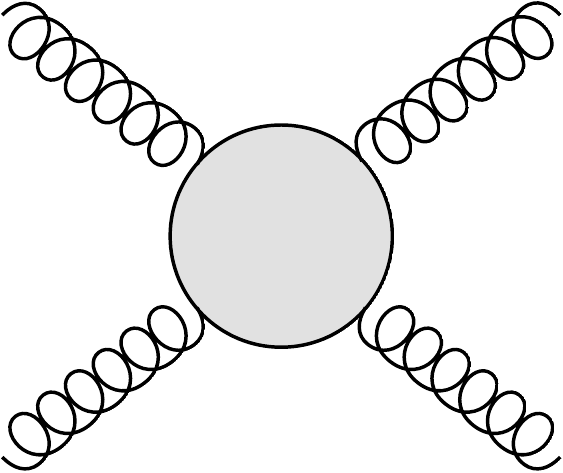}}
 & \parbox{10cm}{\small The four-gluon amplitude has $D=0$ and is logarithmically divergent. The argument holding in QED, stating that gauge invariance renders the four-photon amplitude UV finite, does not apply in QCD, since in a non-abelian gauge theory the elementary four-gluon vertex is part of the gauge-invariant Lagrangian.} \\[9mm]
\parbox{3cm}{\hspace{4.1mm}\includegraphics[scale=0.36]{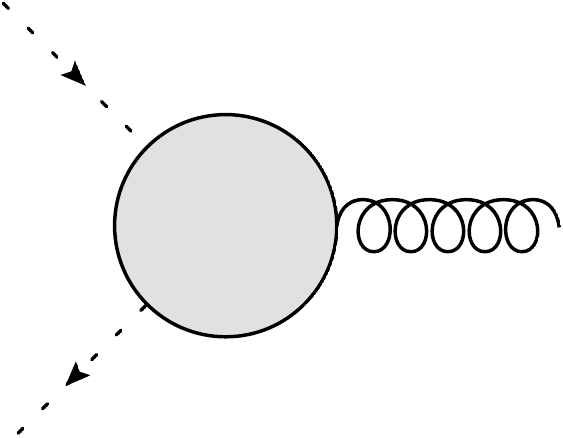}}
 & \parbox{10cm}{\small The ghost--gluon amplitude has $D=0$ and is logarithmically divergent.} \\
\end{array}
\]

In analogy with (\ref{WFR}), we introduce field renormalization constants as
\begin{equation}\label{eq55}
   \psi_{q,0} = Z_2^{1/2}\,\psi_q \,, \qquad
   A_{\mu,0}^a = Z_3^{1/2}\,A_\mu^a \,, \qquad
   c_0^a = Z_{2c}^{1/2}\,c^a \,.
\end{equation}
Since we neglect the light-quark masses, there is no mass renormalization to consider. We must, however, consider the renormalization of the bare QCD coupling constant $g_{s,0}$. Proceeding as in (\ref{meren}), we would define
\begin{equation}
   Z_2\,Z_3^{1/2}\,g_{s,0} = \mu^{\frac{4-d}{2}} Z_1\,g_s \,,
\end{equation}
where $Z_1$ is the renormalization constant associated with the quark--gluon vertex function. However, gauge invariance requires that in QCD {\em all\/} interaction vertices are expressed in terms of the same coupling constant, and this feature must be preserved by renormalization. We can thus express the relation between the bare and renormalized couplings in four different ways, using the quark--gluon, three-gluon, four-gluon, and ghost--gluon vertex functions. This yields the following exact relations between renormalization factors:  
\begin{equation}\label{gs4ren}
\begin{aligned}
   g_{s,0} &= \mu^{\frac{4-d}{2}} Z_1\,Z_2^{-1}\,Z_3^{-1/2}\,g_s \\
   &= \mu^{\frac{4-d}{2}} Z_1^{3g}\,Z_3^{-3/2}\,g_s \\
   &= \mu^{\frac{4-d}{2}} \big(Z_1^{4g}\big)^{1/2}\,Z_3^{-1}\,g_s \\
   &= \mu^{\frac{4-d}{2}} Z_1^{cg}\,Z_{2c}^{-1}\,Z_3^{-1/2}\,g_s \,,
\end{aligned}
\end{equation}
where $Z_1^{3g}$, $Z_1^{4g}$, and $Z_1^{cg}$ denote the renormalization constants associated with the three-gluon, four-gluon, and ghost--gluon vertex functions, respectively. The remaining factors arise when the bare fields entering these vertices are expressed in terms of renormalized fields. Note that, unlike in QED, we no longer have the identity $Z_1=Z_2$ in QCD, since the Ward--Takahashi identity (\ref{Ward}) must be generalized to the more complicated Slavnov-Taylor identities \cite{tHooft:1971akt,Taylor:1971ff,Slavnov:1972fg}. It follows from the above relations that 
\begin{equation}\label{Slavnov}
   Z_1^{3g} = Z_1\,Z_2^{-1}\,Z_3 \,, \qquad
   Z_1^{4g} = \big(Z_1\,Z_2^{-1}\big)^2\,Z_3 \,, \qquad
   Z_1^{cg} = Z_1\,Z_2^{-1}\,Z_{2c} \,.
\end{equation}
These are exact relations between renormalization constants, which hold to all orders in perturbation theory. 

\section{Calculation of the renormalization factors}

The calculation of the renormalization factors $Z_1$, $Z_2$ and $Z_3$ proceeds in analogy to the corresponding calculation in QED. We now briefly summarize the results.

\subsection{Quark self energy}

The calculation of the one-loop quark self energy in the limit of vanishing quark mass is a straightforward application of the loop techniques we have reviewed in Section~\ref{sec:2.2} (problem~2.2). The relevant diagram is:

\begin{center}
\includegraphics[scale=0.5]{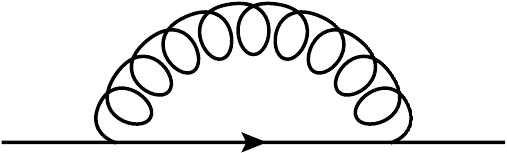}
\end{center}
For the field renormalization constant of the quark field in the $\overline{\rm MS}$ scheme we obtain from~(\ref{basicrenorm})
\begin{equation}\label{Z2QCD}
   Z_2 = 1 - \frac{C_F\alpha_s}{4\pi\hat\epsilon}\,\xi + {\mathcal O}(\alpha_s^2) \,.
\end{equation}
Compared with the corresponding relation (\ref{ZiMSbar}) in QED, where we had worked in Feynman gauge ($\xi=1$), we find a simple replacement $\alpha\to C_F\alpha_s$ accounting for the difference in gauge couplings and the color factor of the one-loop self-energy diagram. 

\subsection{Gluon vacuum polarization}

In addition to the fermion loop graph present in QED, the QCD vacuum polarization function receives several other contributions, which are of genuinely non-abelian origin:
\begin{center}
\includegraphics[scale=0.45]{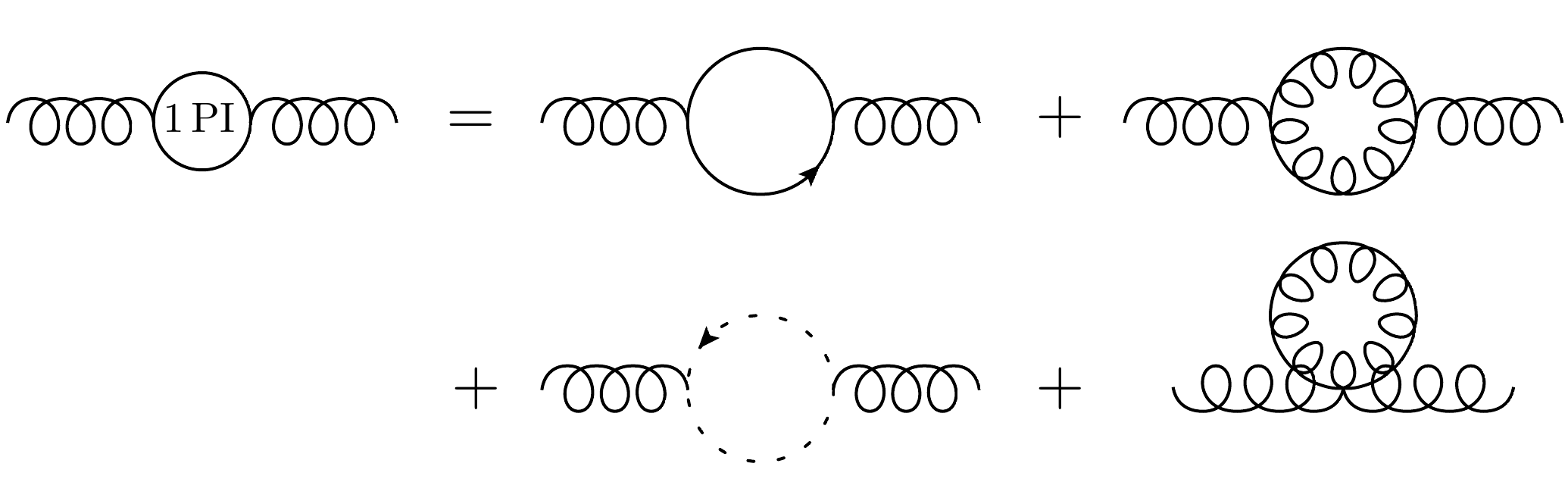}
\end{center}
In analogy with (\ref{pimunu}), we decompose the gluon two-point function in the form
\begin{equation}
   \pi_{\mu\nu}^{ab}(k) = \delta^{ab} \left( k^2 g_{\mu\nu} - k_\mu k_\nu \right) \pi(k^2) \,.
\end{equation}
The fermion loop graph is obtained from the corresponding diagram in QED by means of the replacement $\alpha\to T_F\alpha_s$, where the factor $T_F$ arises from the trace over color matrices. The calculation of the remaining diagrams is a bit more involved. Each individual diagram is quadratically UV divergent, and only a consistent regularization scheme such as dimensional regularization allows one to deal with these divergences in such a way that gauge invariance is preserved. After a lengthy calculation, one obtains
\begin{equation}
   \pi(k^2) = \frac{\alpha_s}{4\pi} \left\{ \left[ \left( \frac{13}{6} - \frac{\xi}{2} \right) C_A
    - \frac43\,T_F\,n_q \right] \left( \frac{1}{\hat\epsilon} - \ln\frac{-k^2-i0}{\mu^2} \right) 
    + \dots \right\} \,,
\end{equation}
where $\xi$ is the gauge parameter and $n_q$ denotes the number of light (approximately massless) quark flavors. For the gluon-field renormalization constant in (\ref{eq55}) we thus obtain the gauge-dependent expression
\begin{equation}
   Z_3 = 1 + \frac{\alpha_s}{4\pi\hat\epsilon} \left[ \left( \frac{13}{6} - \frac{\xi}{2} \right) C_A
    - \frac43\,T_F\,n_q \right] + {\mathcal O}(\alpha_s^2) \,.
\end{equation}

\subsection{Wave-function renormalization for the ghost field}
At one-loop order, the ghost propagator receives the correction:\\
\begin{center}
\includegraphics[scale=0.5]{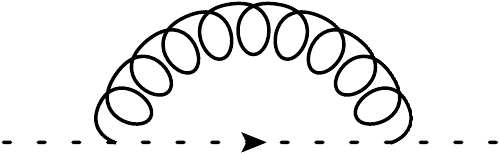}
\end{center}
From a straightforward calculation of this diagram one can extract the wave-function renormalization constant of the ghost field in a general covariant gauge (problem~2.3). The result is
\begin{equation}\label{Z2c}
   Z_{2c} = 1 + \frac{C_A\alpha_s}{4\pi\hat\epsilon}\,\frac{3-\xi}{4} + {\mathcal O}(\alpha_s^2) \,. 
\end{equation}

\subsection{Quark--gluon vertex function}
\label{subsec:3.2.4}

The 1PI one-loop diagrams contributing to the quark--gluon vertex function are:\\[-3mm]

\begin{center}
\includegraphics[scale=0.45]{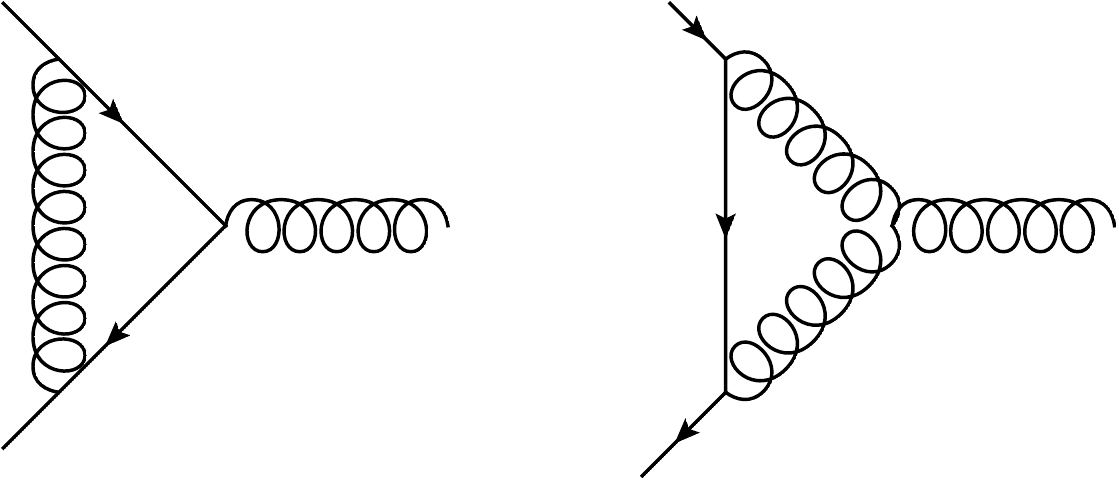}
\end{center}
While the first diagram can be obtained from the corresponding QED diagram by the replacement $\alpha\to C_F\alpha_s$, the second graph is of genuinely non-abelian origin. Its calculation requires the color identity
\begin{equation}
   f_{abc}\,t_b\,t_c = \frac{i}{2}\,C_A\,t_a \,.
\end{equation}
One obtains (problem~2.4)
\begin{equation}
   Z_1 = 1 - \frac{\alpha_s}{4\pi\hat\epsilon} \left( \xi\,C_F + \frac{3+\xi}{4}\,C_A \right)
    + {\mathcal O}(\alpha_s^2) \,.
\end{equation}
Notice that the ``abelian'' part of this result (the term proportional to $C_F$) is the same as in the expression for $Z_2$ in (\ref{Z2QCD}), however the ``non-abelian'' part (the term proportional to $C_A$) violates the identity $Z_1=Z_2$.

\subsection{Charge renormalization}

From the first relation in (\ref{gs4ren}), we now obtain for the charge renormalization constant
\begin{equation}
   Z_g = Z_1\,Z_2^{-1}\,Z_3^{-1/2}
   = 1 - \frac{\alpha_s}{4\pi\hat\epsilon} \left( \frac{11}{6}\,C_A - \frac23\,T_F\,n_q \right)
    + {\mathcal O}(\alpha_s^2) \,.
\end{equation}
Notice that the dependence on the gauge parameter $\xi$ has disappeared. Compared with the corresponding QED relation (where the factor $n_\ell$ counts the number of lepton species)
\begin{equation}
   Z_e = 1 + \frac{\alpha}{6\pi\hat\epsilon}\,n_\ell + {\mathcal O}(\alpha^2) \,,
\end{equation}
one observes that the fermion contributions are identical up to the color factor $T_F=1/2$, while the non-abelian contribution proportional to $C_A$ has no counterpart in QED. Crucially, this contribution has the opposite sign of the fermion contribution \cite{Gross:1973id,Politzer:1973fx}, and this is the reason for the different behavior of the running coupling constants in QED and QCD (see below). 

It is straightforward to calculate the remaining QCD vertex renormalization factors from the relations in (\ref{Slavnov}). We find
\begin{equation}
\begin{aligned}
   Z_1^{3g} &= 1 + \frac{\alpha_s}{4\pi\hat\epsilon} \left[ \left( \frac{17}{12} - \frac{3\xi}{4} \right) C_A
    - \frac43\,T_F\,n_q \right] + {\mathcal O}(\alpha_s^2) \,, \\
   Z_1^{4g} &= 1 + \frac{\alpha_s}{4\pi\hat\epsilon} \left[ \left( \frac{2}{3} - \xi \right) C_A
    - \frac43\,T_F\,n_q \right] + {\mathcal O}(\alpha_s^2) \,, \\
   Z_1^{cg} &= 1 - \frac{\alpha_s}{4\pi\hat\epsilon}\,\frac{\xi}{2}\,C_A + {\mathcal O}(\alpha_s^2) \,. 
\end{aligned}
\end{equation}

\section{Homework problems}

\begin{enumerate}
\item[2.1]
Derive relation (\ref{DQCD}) for the superficial degree of divergence of 1PI QCD Feynman graphs.
\item[2.2]
Calculate the one-loop corrections to the quark self-energy in QCD in the limit of vanishing quark mass. 
\item[2.3]
Derive the expression (\ref{Z2c}) for the ghost-field wave-function renormalization constant $Z_{2c}$ in a general covariant gauge.
\item[2.4]
Calculate the second diagram contributing at one-loop order to the quark--gluon vertex function (see Section~\ref{subsec:3.2.4}) in a general covariant gauge.
\end{enumerate}

\chapter{RG equations and running couplings}
\label{sec:RGEs}

Now that we have discussed the basics of renormalization in both QED and QCD, we will more systematically explore the concept of running couplings and its relevance for multi-scale problems in quantum field theory. A closely related subject is that of the resummation of large logarithmic corrections to all orders of perturbation theory.

Consider a QED observable ${\mathcal O}$ such as a scattering cross section calculated in both the on-shell renormalization scheme and the $\overline{\rm MS}$ scheme. We have
\begin{equation}
   {\mathcal O} = {\mathcal O}_{\rm OS}\bigg( \alpha, m, \ln\frac{s}{m^2}, \dots \bigg) 
   = {\mathcal O}_{\overline{\rm MS}}\bigg( \alpha(\mu), m(\mu), \ln\frac{s}{\mu^2}, \dots \bigg) ,
\end{equation}
where $\sqrt{s}$ is the center-of-mass energy, and the dots refer to other kinematic variables such as scattering angles. In the first expression, $\alpha$ is the fine-structure constant defined in the Thomson limit $q^2\to 0$ and $m$ is the physical mass of the electron, see Section~\ref{sec:2.2}. Both parameters are fundamental physical constants. In the second expression, $\alpha(\mu)$ and $m(\mu)$ are $\mu$-dependent parameters defined in the $\overline{\rm MS}$ scheme. They are related to the parameters in the on-shell scheme via the relations (\ref{MSbarrun}). Several comments are in order:
\begin{itemize}
\item
Both results for the observable ${\mathcal O}$ are equivalent and $\mu$ independent (assuming one works to all orders in perturbation theory, otherwise differences arise only beyond the order to which the calculations have been performed). 
\item
Sometimes on-shell renormalization is inconvenient, because it leaves large logarithmic terms in the expression for the observable. For example, if $s\gg m^2$ for a high-energy process, then $\ln(s/m^2)$ is a large logarithm. Typically, these logarithms appear as $[\alpha\ln(s/m^2)]^n$ in higher orders, and they can threaten the convergence of the perturbative expansion if $\alpha\ln(s/m^2)={\mathcal O}(1)$.\footnote{While for QED this condition would only be satisfied for exceedingly large values of $\sqrt{s}$, large logarithms frequently arise in QCD applications.} 
\item 
Choosing the renormalization scale such that $\mu\approx\sqrt{s}$ fixes this problem, giving a well behaved perturbative expansion in terms of the parameters $\alpha(\mu)$ and $m(\mu)$ with $\mu\approx\sqrt{s}$. These are, however, different from the ``physical'' electron mass and fine-structure constant. As shown in (\ref{MSbarrun}), the corresponding relations are
\begin{equation}
\begin{aligned}
   \alpha(\sqrt{s}) &= \alpha \left( 1 + \frac{\alpha}{3\pi}\,\ln\frac{s}{m^2} + \dots \right) , \\
   m(\sqrt{s}) &= m \left[ 1 - \frac{3\alpha}{4\pi} \left( \ln\frac{s}{m^2} + \frac43 \right) + \dots \right] .
\end{aligned}
\end{equation}
While choosing $\mu\approx\sqrt{s}$ eliminates the large logarithms from the observable itself, it leads to large logarithms in the relations between the parameters in the on-shell scheme and those in the $\overline{\rm MS}$ scheme. We will discuss in a moment how these logarithms can be resummed.
\end{itemize} 

The $\mu$ independence of the observable ${\mathcal O}$ in the $\overline{\rm MS}$ scheme can be expressed in terms of the partial differential equation
\begin{equation}\label{RGE}
   \mu\,\frac{d}{d\mu}\,{\mathcal O}
   = \mu\,\frac{d\alpha(\mu)}{d\mu}\,\frac{\partial{\mathcal O}}{\partial\alpha(\mu)}
    + \mu\,\frac{dm(\mu)}{d\mu}\,\frac{\partial{\mathcal O}}{\partial m(\mu)}
    + \frac{\partial{\mathcal O}}{\partial\ln\mu} = 0 \,.
\end{equation}
Equations of this type are referred to as renormalization-group (RG) equations. They play a fundamental role in the theory of renormalization. The corresponding equations for Green's functions are called Callan--Symanzik equations \cite{Callan:1970yg,Symanzik:1970rt,Symanzik:1971vw}. Above we have assumed that the observable depends on the two running parameters $\alpha(\mu)$ and $m(\mu)$; if it depends on more than two parameters, then (\ref{RGE}) needs to be generalized accordingly.

To proceed, we define two functions of the coupling $\alpha(\mu)$ via
\begin{equation}\label{RGEsQED}
\begin{aligned}
   \mu\,\frac{d\alpha(\mu)}{d\mu} &= \beta\big(\alpha(\mu)\big) \,, \\
   \mu\,\frac{dm(\mu)}{d\mu} &= \gamma_m\big(\alpha(\mu)\big)\,m(\mu) \,.
\end{aligned}
\end{equation}
The first is referred to as the $\beta$-function of QED (admittedly a somewhat dull name), while the second function is called the anomalous dimension of the electron mass. The RG equation (\ref{RGE}) now takes the form
\begin{equation}
   \beta(\alpha)\,\frac{\partial{\mathcal O}}{\partial\alpha}
    + \gamma_m(\alpha)\,m\,\frac{\partial{\mathcal O}}{\partial m}
    + \frac{\partial{\mathcal O}}{\partial\ln\mu} = 0 \,.
\end{equation}
The strategy for obtaining reliable perturbative results in QED, which are free of large logarithms, is now as follows:
\begin{enumerate}
\item
Compute the observable to a given order in perturbation theory in terms of renormalized parameters defined in the $\overline{\rm MS}$ scheme.
\item
Eliminate large logarithms in the expansion coefficients by a suitable choice of the renormalization scale $\mu$.
\item
Compute the running parameters such as $\alpha(\mu)$ and $m(\mu)$ at that scale by solving the differential equations (\ref{RGEsQED}). The boundary values in these solutions can be taken as the fine-structure constant $\alpha$ and the physical electron mass $m$, which are known to excellent accuracy.
\end{enumerate}

The same discussion applies to QCD, where using the $\overline{\rm MS}$ scheme is the default choice. Setting the light quark masses to zero, we obtain the simpler RG equation  
\begin{equation}
   \beta(\alpha_s)\,\frac{\partial{\mathcal O}}{\partial\alpha_s}
    + \frac{\partial{\mathcal O}}{\partial\ln\mu} = 0 \,,
\end{equation}
where
\begin{equation}\label{betadef}
   \beta\big(\alpha_s(\mu)\big) = \mu\,\frac{d\alpha_s(\mu)}{d\mu} \,.
\end{equation}
The running coupling $\alpha_s(\mu)$ is obtained by integrating this equation, using as boundary value the value of $\alpha_s$ at some reference scale, where it is known with high accuracy. A common choice is $\alpha_s(m_Z)=0.1181(11)$ \cite{Tanabashi:2018oca}.

\section{\boldmath Calculation of $\beta$-functions and anomalous dimensions}

There is an elegant formalism that allows us to extract $\beta$-functions and anomalous dimensions from the $1/\epsilon$ poles of the renormalization factors for the various quantities in a quantum field theory. We present the following discussion for the case of QCD with a massive quark of mass $m_q$, but the same results with obvious replacements apply to QED. From the first equation in (\ref{gs4ren}) and the QCD analogue of the first equation in (\ref{meren}), we recall the relations between the bare and renormalized color charge and mass parameter in the form
\begin{equation}
\begin{aligned}
   \alpha_{s,0} &= \mu^{2\epsilon}\,Z_1^2\,Z_2^{-2}\,Z_3^{-1}\,\alpha_s(\mu)
    \equiv \mu^{2\epsilon}\,Z_\alpha(\mu)\,\alpha_s(\mu) \,, \\
   m_{q,0} &= Z_m\,Z_2^{-1}\,m_q(\mu) \equiv Z_m'(\mu)\,m_q(\mu) \,,
\end{aligned}
\end{equation}
where in the $\overline{\rm MS}$ scheme (for QED one replaces $\alpha_s\to\alpha$, $C_F\to 1$ and $\beta_0\to -\frac43\,n_\ell$)
\begin{equation}\label{eq77}
\begin{aligned}
   Z_\alpha(\mu) &= 1 - \beta_0\,\frac{\alpha_s(\mu)}{4\pi\hat\epsilon} + {\mathcal O}(\alpha_s^2) \,;
    \quad \beta_0 = \frac{11}{3}\,C_A - \frac43\,T_F\,n_q \,, \\
   Z_m'(\mu) &= 1 - 3C_F\,\frac{\alpha_s(\mu)}{4\pi\hat\epsilon} + {\mathcal O}(\alpha_s^2) \,.
\end{aligned}
\end{equation}
From the fact that the bare parameters are scale independent it follows that
\begin{equation}
   \mu\,\frac{d}{d\mu}\,\alpha_{s,0} = 0
   = \mu^{2\epsilon}\,Z_\alpha(\mu)\,\alpha_s(\mu) \left[ 2\epsilon + Z_\alpha^{-1}\,\frac{dZ_\alpha}{d\ln\mu}
    + \frac{1}{\alpha_s}\,\frac{d\alpha_s}{d\ln\mu} \right] ,
\end{equation}
which implies
\begin{equation}\label{betadefeps}
   \frac{d\alpha_s}{d\ln\mu} = \alpha_s \left[ - 2\epsilon - Z_\alpha^{-1}\,\frac{dZ_\alpha}{d\ln\mu} \right] 
   \equiv \beta(\alpha_s(\mu),\epsilon) \,,
\end{equation}
and
\begin{equation}
   \frac{d m_{q,0}}{d\ln\mu} = 0
   = Z_m'(\mu)\,m_q(\mu) \left[ Z_m'^{-1}\,\frac{dZ_m'}{d\ln\mu}
    + \frac{1}{m_q}\,\frac{d m_q}{d\ln\mu} \right] ,
\end{equation}
from which it follows that
\begin{equation}\label{mqrun}
   \frac{1}{m_q(\mu)}\,\frac{d m_q(\mu)}{d\ln\mu} 
   = - Z_m'^{-1}\,\frac{dZ_m'}{d\ln\mu}
   \equiv \gamma_m(\alpha_s(\mu)) \,.
\end{equation}
Note that the generalized $\beta$-function $\beta(\alpha_s,\epsilon)$ in (\ref{betadefeps}) governs the scale dependence of the gauge coupling in the regularized theory at finite $\epsilon$. The limit $\epsilon\to 0$ of this expression will later give us the ``usual'' QCD $\beta$-function in the renormalized theory.

We will now derive some beautiful and very useful relations for the $\beta$-function and the anomalous dimension $\gamma_m$. For the purposes of this discussion it is convenient to consider, for a moment, the original MS scheme, in which the $Z$ factors only contain $1/\epsilon^n$ pole terms (with $\epsilon=(4-d)/2$) and thus depend on $\mu$ only through the running coupling $\alpha_s(\mu)$. We can thus write (with $\alpha_s\equiv\alpha_s(\mu)$ throughout)
\begin{equation}\label{eq78}
\begin{aligned}
   \beta(\alpha_s,\epsilon) 
   &= \alpha_s \left[ - 2\epsilon - \beta(\alpha_s,\epsilon)\,Z_\alpha^{-1}\,\frac{dZ_\alpha}{d\alpha_s} \right] , \\
   \gamma_m(\alpha_s) &= - \beta(\alpha_s,\epsilon)\,Z_m'^{-1}\,\frac{dZ_m'}{d\alpha_s} \,.
\end{aligned}
\end{equation}
To solve the first equation one expands 
\begin{equation}
\begin{aligned}
   \beta(\alpha_s,\epsilon) 
   &= \beta(\alpha_s) + \sum_{k=1}^\infty\,\epsilon^k\,\beta^{[k]}(\alpha_s) \,, \\
   Z_\alpha &= 1 + \sum_{k=1}^\infty\,\frac{1}{\epsilon^k}\,Z_\alpha^{[k]}(\alpha_s) \,.
\end{aligned}
\end{equation}
Note that the expansion coefficients $Z_\alpha^{[k]}(\alpha_s)$ start at ${\mathcal O}(\alpha_s^k)$. From the fact that the pole terms $\sim 1/\epsilon^n$ with $n\ge 1$ must cancel in the first relation in (\ref{eq78}) one can derive an infinite set of relations between $\beta^{[k]}(\alpha_s)$ and $Z_\alpha^{[k]}(\alpha_s)$. The solution to this set of equations is (problem~3.1) 
\begin{equation}\label{betamagic}
\begin{aligned}
   \beta^{[1]}(\alpha_s) &= - 2\alpha_s \,, \qquad
    \beta^{[k]}(\alpha_s) = 0 \quad \mbox{for all $k\ge 2$} \,, \\
   \beta(\alpha_s) &= 2\alpha_s^2\,\frac{dZ_\alpha^{[1]}(\alpha_s)}{d\alpha_s} \,.
\end{aligned}
\end{equation}
This yields the exact relation
\begin{equation}
   \beta(\alpha_s,\epsilon) 
   = - 2\epsilon\,\alpha_s + \beta(\alpha_s) = - 2\epsilon\,\alpha_s
    + 2\alpha_s^2\,\frac{dZ_\alpha^{[1]}(\alpha_s)}{d\alpha_s} \,.
\end{equation}
Likewise, one can show that 
\begin{equation}\label{gammamagic}
   \gamma_m(\alpha_s) 
   = 2\alpha_s\,\frac{dZ_m^{\prime[1]}(\alpha_s)}{d\alpha_s} \,.
\end{equation}
Also this result is exact. The above relations state that the $\beta$-function and anomalous dimension can be computed, to all orders in perturbation theory, in terms of the coefficient of the single $1/\epsilon$ pole in the renormalization factors $Z_\alpha$ and $Z_m'$, respectively. Since the coefficients of the $1/\epsilon$ pole terms in the $Z$-factors are the same in the MS and $\overline{\rm MS}$ schemes, these relations also apply to the $\overline{\rm MS}$ scheme. In the one-loop approximation, we obtain from (\ref{eq77}) 
\begin{equation}
   \beta(\alpha_s) = - 2\alpha_s \left( \beta_0\,\frac{\alpha_s}{4\pi} + \dots \right) , \qquad
   \gamma_m(\alpha_s) = - 6 C_F\,\frac{\alpha_s}{4\pi} + \dots \,.
\end{equation}

\section{Leading-order solutions to the evolution equations}
\label{sec:4.2}

\begin{figure}
\begin{center}
\includegraphics[width=0.48\textwidth]{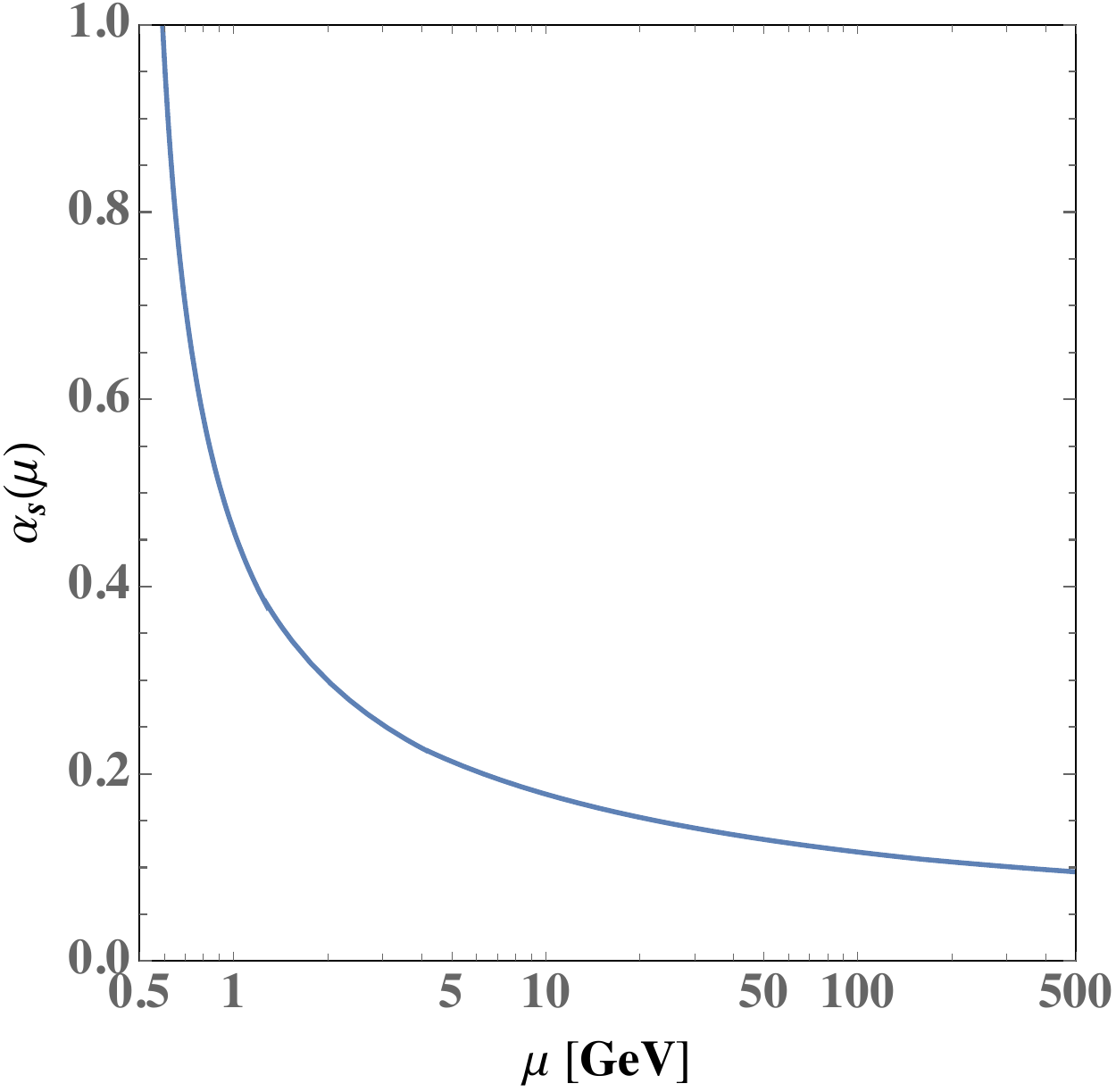} \quad
\includegraphics[width=0.48\textwidth]{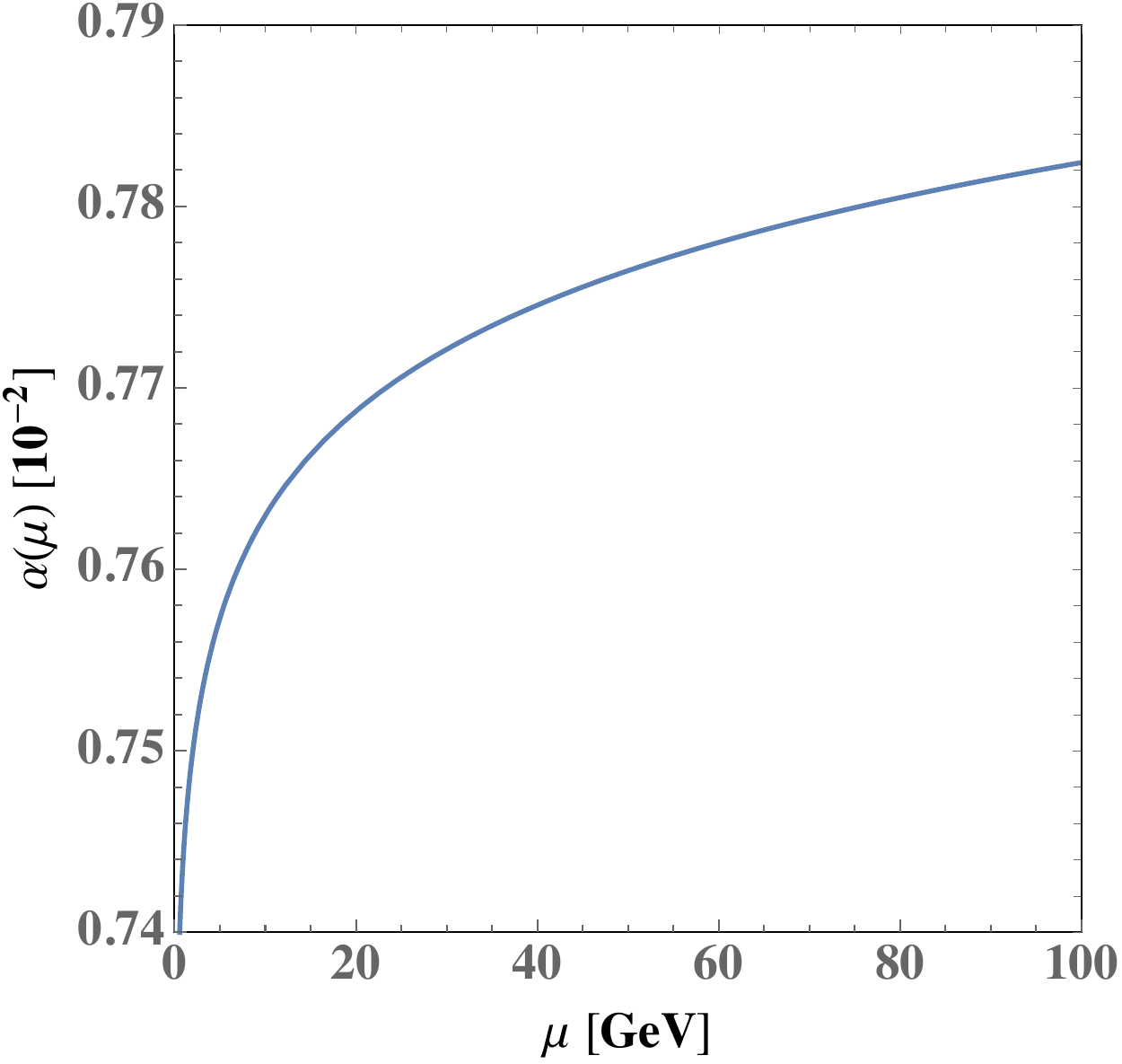}
\caption{\label{fig:running}
Scale dependence of the running QCD coupling $\alpha_s(\mu)$ (left panel) and the running QED coupling $\alpha(\mu)$ (right panel).}
\end{center}
\end{figure}

In the one-loop approximation for the $\beta$-function, equation (\ref{betadef}) governing the scale dependence (also called the ``running'') of the QCD gauge coupling reads 
\begin{equation}
   \frac{d\alpha_s(\mu)}{d\ln\mu} = - 2\beta_0\,\frac{\alpha_s^2(\mu)}{4\pi} \,,
\end{equation}
which using a separation of variables can be rewritten in the form
\begin{equation}
   - \frac{d\alpha_s}{\alpha_s^2} = \frac{\beta_0}{4\pi}\,d\ln\mu^2 \,.
\end{equation}
This can be integrated to obtain
\begin{equation}
   - \int_{\alpha_s(Q)}^{\alpha_s(\mu)}\!\frac{d\alpha_s}{\alpha_s^2} 
   = \frac{1}{\alpha_s(\mu)} - \frac{1}{\alpha_s(Q)}
   = \frac{\beta_0}{4\pi}\,\ln\frac{\mu^2}{Q^2} \,.
\end{equation}
Here $Q$ is some reference scale, at which the value of $\alpha_s(Q)$ is measured with accuracy. A canonical choice is to take $Q$ equal to the mass of the heavy $Z$ boson, $Q=m_Z\approx 91.188$\,GeV, at which $\alpha_s(Q)=0.1181(11)$ \cite{Tanabashi:2018oca}. Rearranging the above result, we find the familiar form of the running coupling in QCD:
\begin{equation}\label{asrun}
   \alpha_s(\mu) = \frac{\alpha_s(Q)}{1+\alpha_s(Q)\,\frac{\beta_0}{4\pi}\,\ln\frac{\mu^2}{Q^2}} \,; \quad
   \beta_0 = \frac{11}{3}\,C_A - \frac43\,T_F\,n_q \,.
\end{equation}
Here $n_q$ is the number of light (massless) quark flavors with masses below the scale $\mu$. The corresponding expression for QED is obtained by replacing $\alpha_s\to\alpha$ and $\beta_0\to-\frac43\,n_\ell$. Figure~\ref{fig:running} shows the two couplings as a function of the energy scale. It is not difficult to include higher-order corrections in the calculation of the running couplings of QCD and QED, see Section~\ref{sec:5.7}. These higher-order corrections are included in the figure. 

Note that $\beta_0$ is {\em positive\/} in QCD (since the number of quark generations is less than 17), while it is negative in QED. As a result, the strong coupling gets weaker at higher energies -- a phenomenon referred to as {\em asymptotic freedom\/} \cite{Gross:1973id,Politzer:1973fx}, which was awarded the 2004 Nobel Prize in Physics -- while the QED coupling slowly increases with energy. At low energies the QCD coupling grows, and the leading-order expression (\ref{asrun}) blows up at the scale
\begin{equation}\label{eq92}
   \mu = Q\,\exp\bigg(-\frac{2\pi}{\beta_0\,\alpha_s(Q)}\bigg) \approx 0.2\,\mbox{GeV} \,.
\end{equation}
QCD becomes strongly coupled at such low scales, and the quarks and gluons are confined inside hadrons. The chiral Lagrangian provides an effective theory for QCD at such low scales. This is discussed in the lectures by Antonio Pich \cite{Pich:2018ltt} elsewhere in this book. For QED, the evolution effects of the gauge coupling are more modest but not negligible. At $\mu=m_Z$, the value of $\alpha(m_Z)$ is about 6\% larger than the fine-structure constant, which according to (\ref{MSbarrun}) corresponds to the $\overline{\rm MS}$ coupling evaluated at the scale $\mu=m\approx 511$\,keV. At very high values of $\mu$, the QED running coupling develops a so-called Landau pole and diverges.

Note that, in the $\overline{\rm MS}$ scheme, the slope of the running coupling changes whenever $\mu$ crosses the mass scale of a fermion. The simple form shown in (\ref{asrun}) holds only in an interval where the value of $n_q$ is fixed. When $\mu$ crosses a quark threshold, the value of $\beta_0$ changes abruptly, and the values of $\alpha_s(\mu)$ just above and below the threshold must be matched to each other. For example, at $\mu=m_Z\approx 91.188$\,GeV QCD contains 5 approximately massless quark flavors, while the top quark with mass $m_t\approx 170$\,GeV is heavy and is neglected in the running of the coupling. Formula (\ref{asrun}) can be used to evolve the coupling down to the scale $\mu=m_b(m_b)\approx 4.18$\,GeV, below which the mass of the bottom quark can no longer be neglected. In the $\overline{\rm MS}$ scheme one computes $\alpha_s(m_b)$ from (\ref{asrun}) using $\beta_0=\frac{23}{3}$ (corresponding to $n_q=5$), but for lower scales one replaces (\ref{asrun}) with the analogous relation
\begin{equation}\label{eq93}
   \alpha_s(\mu) = \frac{\alpha_s(m_b)}{1+\alpha_s(m_b)\,\frac{\beta_0}{4\pi}\,\ln\frac{\mu^2}{m_b^2}} \,; \quad
   \mu < m_b \,,
\end{equation}
where now $\beta_0=\frac{25}{3}$ (corresponding to $n_q=4$). The same procedure is repeated when $\mu$ falls below the scale of the charm quark ($m_c(m_c)\approx 1.275$\,GeV), or when $\mu$ is raised above the scale of the top-quark mass. This is explored in more detail in problem~3.2.

Let us now study the scale evolution of the running quark masses in QCD. This is important, since free quarks do not exist due to confinement, so unlike in QED the quark masses must always be defined as running parameters. We can rewrite the evolution equation in (\ref{mqrun}) in the form
\begin{equation}
   \frac{d m_q(\mu)}{d\ln\mu} = \beta(\alpha_s)\,\frac{d m_q(\mu)}{d\ln\alpha_s}
   = m_q(\mu)\,\gamma_m(\alpha_s) \,,
\end{equation}
which using  separation of variables can be recast as
\begin{equation}\label{eq96}
   \frac{d m_q}{m_q} = \frac{\gamma_m(\alpha_s)}{\beta(\alpha_s)}\,d\alpha_s 
   \approx - \frac{\gamma_m^0}{2\beta_0}\,\frac{d\alpha_s}{\alpha_s} \,.
\end{equation}
Here we have expanded the anomalous dimension in the perturbative series
\begin{equation}\label{gammser}
   \gamma_m(\alpha_s) = \gamma_m^0\,\frac{\alpha_s}{4\pi}
    + \gamma_m^1 \left( \frac{\alpha_s}{4\pi} \right)^2 + \dots \,; \quad
   \gamma_m^0 = - 6 C_F 
\end{equation}
and kept the leading term only. Integrating relation (\ref{eq96}) in the leading-order approximation yields
\begin{equation}
   \ln\frac{m_q(\mu)}{m_q(Q)} = - \frac{\gamma_m^0}{2\beta_0}\,\ln\frac{\alpha_s(\mu)}{\alpha_s(Q)} \,,
\end{equation}
and hence
\begin{equation}\label{mqrunresu}
   m_q(\mu) = m_q(Q) \left( \frac{\alpha_s(\mu)}{\alpha_s(Q)} \right)^{- \frac{\gamma_m^0}{2\beta_0}} .
\end{equation}
Since the exponent is positive, it follows that quarks get lighter at higher energies. Again it would not be difficult to include higher-order corrections in this analysis (problem~3.3). 

As an example of this effect, let us study the evolution of the bottom-quark mass from the scale $\mu=m_b$ to the mass scale of the Higgs boson. The resulting parameter $m_b(m_h)$ governs the effective coupling of the Higgs boson to a pair of $b$ quarks. Starting from $m_b(m_b)\approx 4.18$\,GeV, we obtain
\begin{equation}
   m_b(m_h) \approx m_b(m_b) \left( \frac{\alpha_s(m_h)}{\alpha_s(m_b)} \right)^{12/23}
   \approx 2.79\,\mbox{GeV} \,.
\end{equation}
Obviously, evolution effects have a large impact in this case, and ignoring them would largely overestimate the Higgs--bottom coupling at high energies.

\section{Fixed points of running couplings}

Now that we have discussed the concept of running couplings and $\beta$-functions, let me take a moment to talk about fixed points of RG flows. An interesting possibility is that the $\beta$-function $\beta(g)$ for some coupling $g(\mu)$ in a quantum field theory has a zero at some value $g_\star\ne 0$ of the coupling, e.g.:\\
\begin{center}
\includegraphics[scale=0.45]{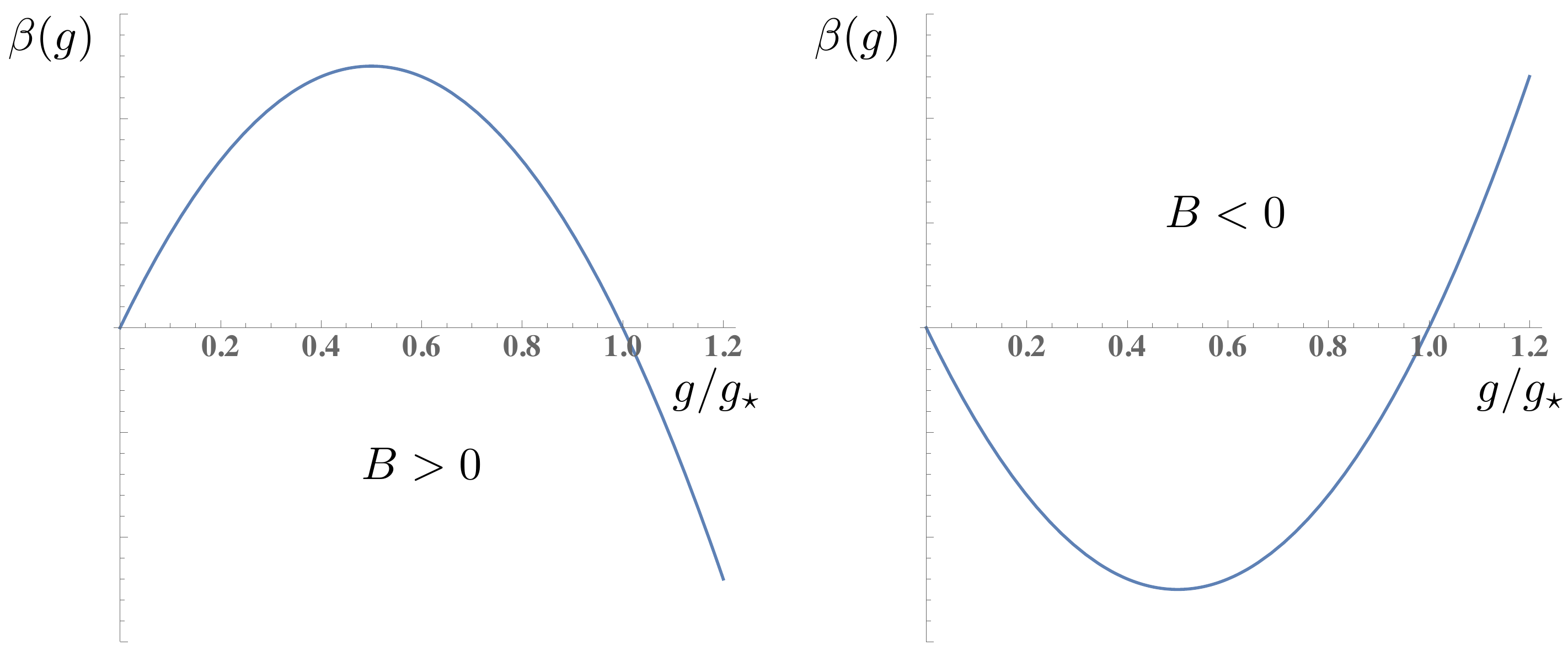}
\vspace{2mm}
\end{center}
Near such a fixed point we have
\begin{equation}
   \beta(g) \approx -B \left( g - g_\star \right) = \frac{dg}{d\ln\mu} \,,
\end{equation}
and integrating this equation yields
\begin{equation}
   g(\mu) \approx g_\star + \left[ g(Q) - g_\star \right] \left( \frac{\mu}{Q} \right)^{-B} .
\end{equation}

We can now distinguish two cases:
\begin{equation}
\begin{aligned}
   B>0: \qquad & g(\mu) \to g_\star \quad \mbox{for $\mu\to\infty$ \quad (UV fixed point)} \\
   B<0: \qquad & g(\mu) \to g_\star \quad \mbox{for $\mu\to 0$ \hspace{5mm} (IR fixed point)} 
\end{aligned}
\end{equation}
Green's functions obey power-like scaling laws near the fixed point, with {\em critical exponents\/} given in terms of anomalous dimensions $\gamma(g_\star)$. Critical phenomena in condensed-matter physics (e.g.\ phase transitions) are described by anomalous dimensions in simple quantum field theories, such as scalar $\phi^4$ theory (see e.g.\ chapters~12 and 13 in \cite{Peskin:1995ev}).

\newpage
\section{Homework problems}

\begin{enumerate}
\item[3.1]
Prove the relations given in (\ref{betamagic}) and (\ref{gammamagic}). The trick is to consider the products $Z_\alpha\,\beta(\alpha_s,\epsilon)$ and $Z_m'\,\gamma_m(\alpha_s)$.
\item[3.2]
In Section~\ref{sec:4.2}, I have described the matching procedure, which needs to be applied in the $\overline{\rm MS}$ scheme whenever the scale $\mu$ in the running coupling $\alpha_s(\mu)$ in (\ref{asrun}) crosses a quark threshold. Using this procedure, computed the values of $\alpha_s(m_t)$, $\alpha_s(m_b)$, and $\alpha_s(m_c)$ starting from $\alpha_s(m_Z)=0.1181$ and using the masses $m_t(m_t)=163.4$\,GeV, $m_b(m_b)=4.18$\,GeV and $m_c(m_c)=1.275$\,GeV, where $m_q(m_q)$ are the running quark masses in the $\overline{\rm MS}$ scheme evaluated at $\mu=m_q$. Also determine the value of $\mu$ at which the leading-order formula for the running coupling blows up, see (\ref{eq92}). Then repeat the same exercise for the running QED coupling. Starting from $\alpha(m_e)=1/137.036$, compute $\alpha(m_\mu)$, $\alpha(m_\tau)$ and $\alpha(m_Z)$.
\item[3.3]
Integrate the differential equation (\ref{eq96}) for the running quark mass in QCD keeping the two-loop coefficients $\beta_1$ and $\gamma_m^1$ in (\ref{betafunc}) and (\ref{gammser}). Expand the ratio $\gamma_m(\alpha_s)/\beta(\alpha_s)$ in $\alpha_s$ to subleading order and integrate the resulting expression. Find the appropriate generalization of (\ref{mqrunresu}), which should be of the form
\[
   m_q(\mu) = m_q(Q) \left( \frac{\alpha_s(\mu)}{\alpha_s(Q)} \right)^{- \frac{\gamma_m^0}{2\beta_0}} 
   \left[ 1 + c_m\,\frac{\alpha_s(\mu)-\alpha_s(Q)}{4\pi} + \dots \right]
\]
with a constant coefficient $c_m$ that you must determine.
\end{enumerate}

\chapter{Effective field theories, composite operators and the Wilsonian approach to renormalization}

The basic idea underlying the construction of an effective field theory is that, in a situation where one is faced with a quantum field theory with two (or more) very different energy or length scales, one can construct a simpler theory by performing a systematic expansion in the ratio of these scales. Let us consider an illustrative example.

In view of the fact that the Standard Model of particles physics leaves many questions unanswered, it is plausible that there should exist some ``physics beyond the Standard Model'' involving new heavy particles with masses $M\gg v$ much above the scale of electroweak symmetry breaking. While the complete Lagrangian of the UV theory is at present still out of sight, we can construct its low-energy effective theory -- the so-called SMEFT -- by extending the familiar Standard Model Lagrangian with higher-dimensional local operators built out of Standard Model fields \cite{Weinberg:1979sa,Wilczek:1979hc,Buchmuller:1985jz,Leung:1984ni,Grzadkowski:2010es}:
\begin{equation}\label{SMEFT}
   {\mathcal L}_{\rm SMEFT} 
   = {\mathcal L}_{\rm SM} + \sum_{n\ge 1} \sum_i \frac{C_i^{(n)}}{M^n}\,{\mathcal O}_i^{(n)} .
\end{equation}
The new operators ${\mathcal O}_i^{(n)}$ with mass dimension $D=4+n$ must respect the symmetries of the Standard Model, such as Lorentz invariance and gauge invariance. There is of course an infinite set of such operators, but importantly there exists only a finite set of operators for each dimension $D$, and the contributions of these operators to any given observable are suppressed by powers of $(v/M)^{D-4}$ relative to the contributions of the operators of the Standard Model. This is discussed in detail in the lectures by Aneesh Manohar \cite{Manohar:2018aog} elsewhere in this book. 

Note that this estimate of the scaling of the higher-order terms assumes that the relevant energies in the process of interest are of order the weak scale $v$. If one considers high-energy processes characterized by an energy $E\gg v$, then  the minimum suppression factor is $(E/M)^{D-4}$ rather than $(v/M)^{D-4}$. An example are transverse-momentum distributions of Standard Model particles produced at the LHC in the region where $p_T\gg v$. If the characteristic energies $E$ are of order the new-physics scale $M$, then the effective field theory in (\ref{SMEFT}) breaks down. Even in this case not all is lost. A different construction based on soft-collinear effective theory \cite{Bauer:2000yr,Bauer:2001ct,Bauer:2001yt,Beneke:2002ph} -- a non-local effective field theory discussed in the lectures by Thomas Becher \cite{Becher:2018gno} elsewhere in this book -- can deal with the case where some kinematical variables in the low-energy theory are parametrically larger than the weak scale \cite{Alte:2018nbn}.
  
Let me briefly recall how an effective Lagrangian such as (\ref{SMEFT}) is derived. ``Integrating out'' the heavy degrees of freedom associated with the high scale $M$ from the generating functional of Green's functions one obtains a non-local action functional, which can be expanded in an infinite tower of {\em local operators} ${\mathcal O}_i^{(n)}$ \cite{Polchinski:1992ed}. For fixed $n$, the $\{{\mathcal O}_i^{(n)}\}$ form a complete set (a basis) of local, $D=4+n$ composite operators built out of the fields of the low-energy theory. These operators are only constrained by the symmetries of the low-energy theory, such as Lorentz invariance, gauge invariance, and global symmetries such as $C$, $P$, $T$, flavor symmetries, etc. The {\em Wilson coefficients\/} $C_i^{(n)}$ are dimensionless (this can always be arranged) and contain all information about the short-distance physics which has been integrated out. The above equation is useful only because the infinite sum over $n$ can be truncated at some value $n_{\rm max}$, since matrix elements of the operators ${\mathcal O}_i^{(n)}$ scale like powers of $m$, where $m\ll M$ represents the characteristic scale of the low-energy effective theory ($m=v$ in the example of SMEFT), i.e.\
\begin{equation}
   \langle f|\,{\mathcal O}_i^{(n)}\,|i\rangle \sim m^{n+\delta} .
\end{equation}
Here $\delta$ is set by the external states. Truncating the sum at $n_{\rm max}$ one makes an error of order $(m/M)^n\ll 1$ relative to the leading term.

\section{Running couplings and composite operators}

\begin{figure}
\begin{center}
\includegraphics[width=0.7\textwidth]{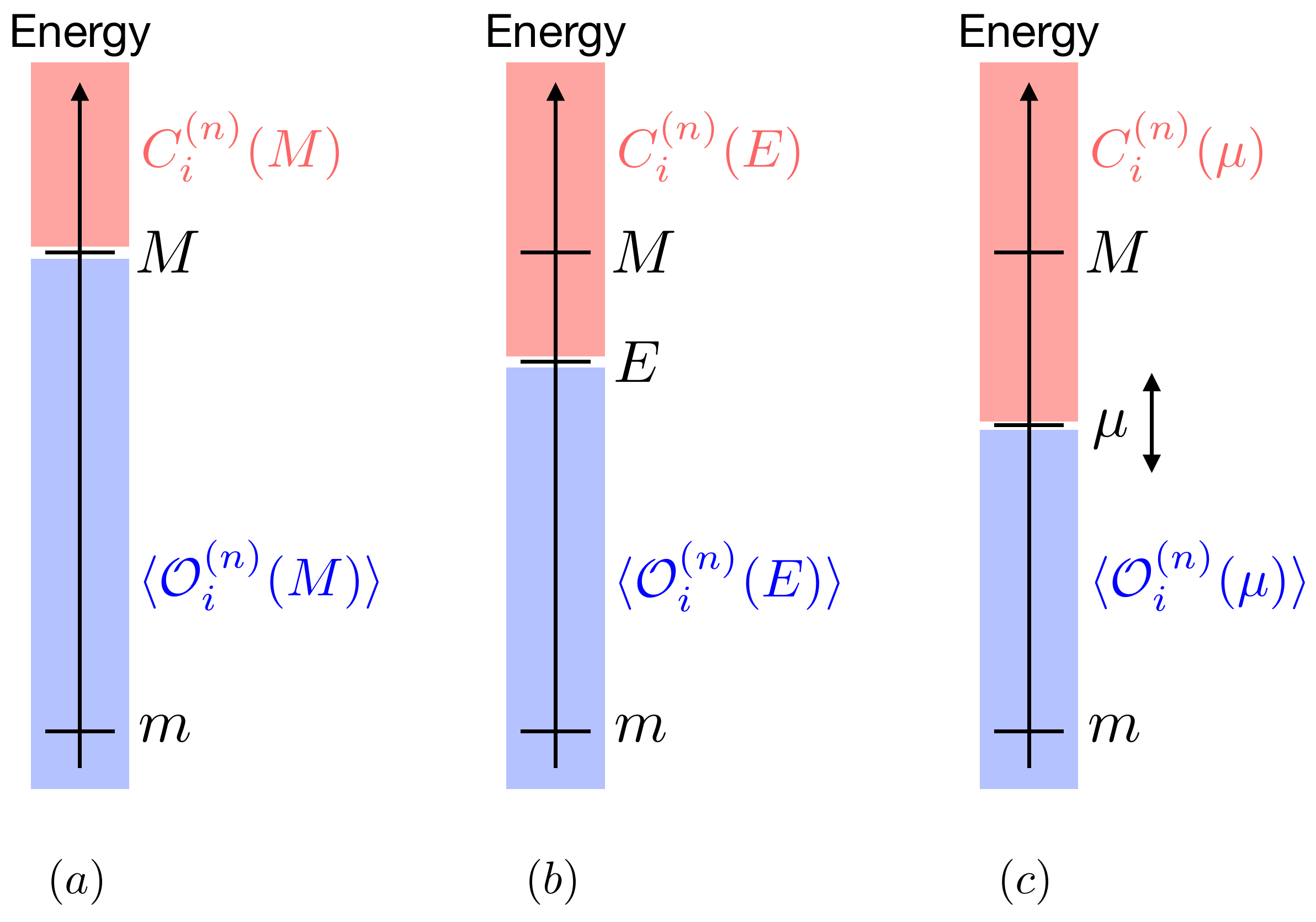}
\caption{\label{fig:scales}
Factorization of an observable into short-distance (red) and long-distance (blue) contributions, which are accounted for by the Wilson coefficients and operator matrix elements of the effective field theory. The panels differ by the choice of the factorization scale.}
\vspace{3mm}
\end{center}
\end{figure}

In essence, in constructing the effective Lagrangian (\ref{SMEFT}) we split up the contributions from virtual particles into short- and long-distance modes:
\begin{equation}
   \int_0^\infty\!\frac{d\omega}{\omega} 
   =  \int_M^\infty\!\frac{d\omega}{\omega} + \int_0^M\!\frac{d\omega}{\omega} \,,
\end{equation}
where the first term is sensitive to UV physics and is absorbed into the Wilson coefficients $C_i^{(n)}$, while the second term is sensitive to IR physics and is absorbed into the matrix elements $\langle{\mathcal O}_i^{(n)}\rangle$. This is illustrated in panel (a) of Figure~\ref{fig:scales}. Now imagine that we are performing a measurement at a characteristic energy scale $E$, such that $m\ll E<M$. We can then integrate out the high-energy fluctuations of the light Standard Model fields (with frequencies $\omega>E$) from the generating functional, because they will not be needed as source terms for external states. This yields a {\em different\/} effective Lagrangian, but one in which the operators ${\mathcal O}_i^{(n)}$ are the same as before (since we have not removed any Standard Model particles). What changes is the split-up of modes, which now reads
\begin{equation}
   \int_0^\infty\!\frac{d\omega}{\omega} 
   =  \int_E^\infty\!\frac{d\omega}{\omega} + \int_0^E\!\frac{d\omega}{\omega} \,,
\end{equation}
as shown in panel (b) of Figure~\ref{fig:scales}. As a consequence, the values of the Wilson coefficients and operators matrix elements need to be different, i.e.\footnote{The terms with $n=0$ account for the ``renormalizable'' (in an old-fashioned sense) interactions, such as the Standard Model Lagrangian in (\ref{SMEFT}).}
\begin{equation}
   {\mathcal L}_{\rm EFT} 
   = \sum_{n=0}^\infty \sum_i \frac{C_i^{(n)}(M)}{M^n}\,{\mathcal O}_i^{(n)}(M)
   = \sum_{n=0}^\infty \sum_i \frac{C_i^{(n)}(E)}{M^n}\,{\mathcal O}_i^{(n)}(E) \,.
\end{equation}
We are thus led to study the effective Lagrangian 
\begin{equation}\label{EFTmu}
   {\mathcal L}_{\rm EFT} 
   = \sum_{n=0}^\infty \sum_i \frac{C_i^{(n)}(\mu)}{M^n}\,{\mathcal O}_i^{(n)}(\mu) \,,
\end{equation}
whose matrix elements are, by construction, independent of the arbitrary factorization scale $\mu$ (with $m\le\mu\le M$), see panel (c) in the figure. Here ${\mathcal O}_i^{(n)}(\mu)$ are {\rm renormalized composite operators\/} defined in dimensional regularization and the $\overline{\rm MS}$ scheme, while $C_i^{(n)}(\mu)$ are the corresponding renormalized Wilson coefficients. These are nothing but the running couplings of the effective theory, in generalization to our discussion of running gauge couplings and running mass parameters in the previous sections. The scale $\mu$ serves as the renormalization scale for these quantities, but at the same time it is the factorization scale which separates short-distance (high-energy) from long-distance (low-energy) contributions.

Several comments are in order:
\begin{itemize}
\item
The terms with $n=0$ are just the renormalizable Lagrangians of the low-energy theory. As a consequence, parameters such as $\alpha_s(\mu)$ or $m(\mu)$  might, in fact, contain some information about short-distance physics through their scale dependence. For example, the $\mu$ dependence of gauge couplings and running mass parameters in supersymmetric extensions of the Standard Model hint at a grand unification of the strong and electroweak forces at a scale $M\sim 10^{16}$\,GeV (see e.g.\ \cite{Ross:1985ai}). 
\item
The higher-dimensional operators with $n\ge 1$ are interesting, because their coefficients tell us something about the fundamental high-energy scale $M$. The most prominent example is that of the weak interactions at low energy. These are described by four-fermion operators with mass dimension $D=6$, whose coefficients are proportional to the Fermi constant $\sqrt{2}\,G_F=1/v^2$ (see \cite{Buchalla:1995vs} for an excellent review). The numerical value of $G_F$ indicates the fundamental mass scale of electroweak symmetry breaking. Indeed, the masses of the heavy weak gauge bosons $W^\pm $ and $Z^0$ could be estimated based on observation of weak decays at low energies, long before these particles were discovered. \item
The Standard Model also contains the dimension-2 operator $\mu^2\phi^\dagger\phi$ (here $\mu^2$ is the Higgs mass parameter, not the renormalization scale), which corresponds to $n=-2$. It follows from the assumption of naturalness that $\mu^2\sim M^2$, and the fact that empirically $|\mu^2|$ is much less than the new-physics scale is known as the {\em hierarchy problem\/} of the Standard Model. In a natural theory of fundamental physics, dimension-2 operators should be forbidden by some symmetry, such as supersymmetry (see e.g.\ \cite{Wess:1992cp}). 
\end{itemize}

At any fixed $n$, the basis $\{{\mathcal O}_i^{(n)}\}$ of composite operators can be renormalized in the standard way, allowing however for the possibility of {\em operator mixing}. In analogy with the corresponding relations for the field operators in (\ref{WFR}) and (\ref{eq55}), we write
\begin{equation}\label{eq109}
   {\mathcal O}_{i,0}^{(n)} = \sum_j Z_{ij}^{(n)}(\mu)\,{\mathcal O}_j^{(n)}(\mu) \,.
\end{equation}
The operators on the left-hand side are bare operators (as denoted by the subscript ``0''), while the operators on the right-hand side are the renormalized operators. In the presence of operator mixing, in the renormalization of the bare operator ${\mathcal O}_{i,0}^{(n)}$ one needs other operators ${\mathcal O}_j^{(n)}(\mu)$ with $j\ne i$ as counterterms. Note that the renormalization constants $Z_{ij}^{(n)}$ contain a wave-function renormalization factor $Z_a^{1/2}$ for each component field contained in the composite operators in addition to renormalization factors absorbing the UV divergences of the 1PI loop corrections to the operator matrix elements. Importantly, in dimensional regularization there is no mixing between operators of different dimension. This fact singles out dimensional regularization as the most convenient regularization scheme. In order not to clutter the notation too much, we will from now on drop the superscript ``$(n)$'' on the operators, their Wilson coefficients and the renormalization factors. 

There are some important facts about the renormalization properties of composite operators, which are discussed for instance in Section~V of \cite{Pascual:1984zb}. One distinguishes three types of composite operators:
\begin{itemize}
\item
Class-I operators are gauge invariant and do not vanish by virtue of the classical equations of motion.
\item
Class-II operators are gauge invariant but their matrix elements vanish by virtue of the classical equations of motion.
\item
Class-III operators are not gauge invariant.
\end{itemize}
In the renormalization of composite operators it is convenient to use the background-field method \cite{Abbott:1980hw}, which offers an elegant method for renormalizing gauge theories while preserving explicit gauge invariance. Then the following statements hold:
\begin{enumerate}
\item
The renormalization of class-I operators involves class-I and class-II operators, but not class-III operators as counterterms. In matrix notation
\begin{equation}\label{classI}
   \vec{\mathcal O}_{\rm I,0} = \bm{Z}_{\rm I}\,\vec{\mathcal O}_{\rm I} + \bm{Z}_{\rm I\to II}\,\vec{\mathcal O}_{\rm II} \,.
\end{equation}
\item
Class-II and class-III operators are renormalized among themselves, i.e.\
\begin{equation}\label{class2ops}
   \vec{\mathcal O}_{\rm II,0} = \bm{Z}_{\rm II}\,\vec{\mathcal O}_{\rm II} \,, \qquad
   \vec{\mathcal O}_{\rm III,0} = \bm{Z}_{\rm III}\,\vec{\mathcal O}_{\rm III} \,.
\end{equation}
\item
Since on-shell matrix elements of class-II operators vanish by the equations of motion, the contribution of class-II operators in (\ref{classI}) has no physical consequences. Furthermore, in background-field gauge class-III operators never arise. Importantly, class-I operators do not appear in (\ref{class2ops}), and hence class-II operators can be ignored for all practical purposes.
\end{enumerate}

Let me add an important comment here. It is often stated that the use of the classical equations of motion to eliminate operators from the basis $\{{\mathcal O}_i^{(n)}\}$ is not justified beyond tree level. This statement is false! Class-II operators can always be removed using {\em field redefinitions}, which corresponds to a change of variables in the functional integral \cite{Politzer:1980me,Georgi:1991ch}.\footnote{In some cases this generates class-I operators of higher dimension.} 
An explicit proof of this statement is presented in Aneesh Manohar's lectures \cite{Manohar:2018aog} elsewhere in this book. Special care must be taken when these field redefinitions change the measure of the functional integral. This happens for the case of a fermionic chiral transformation and gives rise to the famous chiral anomaly \cite{Fujikawa:1979ay}. In any event, the lesson is that at fixed $n\ge 1$ class-II operators can simply be removed from the operator basis. 

\section{Anomalous dimensions of composite operators}

From the fact that the bare operators on the left-hand side of (\ref{eq109}) are scale independent, it follows that (a sum over repeated indices is implied)
\begin{equation}
   \frac{dZ_{ij}(\mu)}{d\ln\mu}\,{\mathcal O}_j(\mu) + Z_{ij}(\mu)\,\frac{d{\mathcal O}_j(\mu)}{d\ln\mu} = 0 \,, 
\end{equation}
which can be solved to give
\begin{equation}
   \frac{d{\mathcal O}_k(\mu)}{d\ln\mu} = - (Z^{-1})_{ki}(\mu)\,\frac{dZ_{ij}(\mu)}{d\ln\mu}\,{\mathcal O}_j(\mu)
   \equiv - \gamma_{kj}(\mu)\,{\mathcal O}_j(\mu) \,.
\end{equation}
In matrix notation, this becomes
\begin{equation}
   \frac{d\vec{{\mathcal O}}(\mu)}{d\ln\mu} = - \bm{\gamma}(\mu)\,\vec{{\mathcal O}}(\mu) \,,
    \quad \mbox{with} \quad
   \bm{\gamma}(\mu) = \bm{Z}^{-1}(\mu)\,\frac{d\bm{Z}(\mu)}{d\ln\mu} \,.
\end{equation}
The quantity $\bm{\gamma}$ is called the {\em anomalous-dimension matrix\/} of the composite operators. In analogy with (\ref{gammamagic}), this quantity can be obtained from the coefficient of the $1/\epsilon$ pole term in $\bm{Z}$ via the exact relation (problem~4.1)
\begin{equation}\label{moremagic}
   \bm{\gamma} = - 2\alpha_s\,\frac{\partial\bm{Z}^{[1]}}{\partial\alpha_s} \,.
\end{equation}
To calculate the anomalous-dimension matrix one first computes the matrix of renormalization factors $\bm{Z}$ in (\ref{eq109}) and then obtains $\bm{\gamma}$ from the coefficient of the single $1/\epsilon$ pole terms.

\section{RG evolution equation for the Wilson coefficients}

The fact that the effective Lagrangian is $\mu$ independent by construction implies, for fixed $n\ge 0$, that (a sum over repeated indices is implied)
\begin{equation}
   \frac{dC_i(\mu)}{d\ln\mu}\,{\mathcal O}_i(\mu) + C_i(\mu)\,\frac{d{\mathcal O}_i(\mu)}{d\ln\mu} 
   = \left[ \frac{dC_i(\mu)}{d\ln\mu}\,\delta_{ij} - C_i(\mu)\,\gamma_{ij}(\mu) \right] {\mathcal O}_j(\mu) = 0 \,.
\end{equation}
From the linear independence of the basis operators, it follows that
\begin{equation}
   \frac{dC_j(\mu)}{d\ln\mu} - C_i(\mu)\,\gamma_{ij}(\mu) = 0
\end{equation}
for each $j$, which in matrix notation can be written as
\begin{equation}\label{eq118}
   \frac{d\vec{C}(\mu)}{d\ln\mu} = \bm{\gamma}^T(\mu)\,\vec{C}(\mu) \,.
\end{equation}
This matrix differential equation governs the RG evolution of the Wilson coefficients.

In order to solve this equation, we first change variables and express the scale dependence of the various objects via the running QCD coupling $\alpha_s(\mu)$. Using (\ref{betadef}), this leads to
\begin{equation}\label{eq119}
   \frac{d\vec{C}(\alpha_s)}{d\alpha_s} 
   = \frac{\bm{\gamma}^T(\alpha_s)}{\beta(\alpha_s)}\,\vec{C}(\alpha_s) \,.
\end{equation}
Apart from a factor of $i$ on the left-hand side, this equation has the same structure as the time-dependent Schr\"odinger equation in quantum mechanics, where in our case $\alpha_s$ plays the role of time, $\vec{C}$ corresponds to the Schr\"odinger wave function, and $\bm{\gamma}^T/\beta$ plays the role of the Hamiltonian. It follows that the general solution of (\ref{eq119}) is
\begin{equation}\label{eq120}
   \vec{C}\big(\alpha_s(\mu)\big) = \mbox{T}_{\alpha_s}\,\exp\left[\,
    \int\limits_{\alpha_s(M)}^{\alpha_s(\mu)}\!d\alpha_s\,\frac{\bm{\gamma}^T(\alpha_s)}{\beta(\alpha_s)}
    \right] \vec{C}\big(\alpha_s(M)\big) \,,
\end{equation}
where the symbol ``$\mbox{T}_{\alpha_s}$'' implies an ordering of the matrix exponential such that matrices are ordered from left to right according to decreasing $\alpha_s$ values, assuming $\alpha_s(\mu)>\alpha_s(M)$. This is the analogue of the time-ordered exponential in the quantum-mechanical expression for the time-evolution operator. The boundary coefficients $\vec{C}\big(\alpha_s(M)\big)$ correspond to the Wilson coefficients at the high matching scale, which can be computed order by order in QCD perturbation theory. The matrix exponential has the effect of evolving (``running'') these coefficients down to a factorization scale $\mu<M$. As we will show in a moment, in this process large logarithms arise (for $\mu\ll M$), which are resummed automatically in the solution (\ref{eq120}). 

At leading order (but not beyond) the ordering symbol becomes irrelevant, and expanding
\begin{equation}\label{betafunc}
   \bm{\gamma}(\alpha_s) = \bm{\gamma}_0\,\frac{\alpha_s}{4\pi}
    + \bm{\gamma}_1 \left( \frac{\alpha_s}{4\pi} \right)^2 + \dots \,, \qquad
   \beta(\alpha_s) = - 2\alpha_s \left[ \beta_0\,\frac{\alpha_s}{4\pi} 
    + \beta_1 \left( \frac{\alpha_s}{4\pi} \right)^2 + \dots \right]
\end{equation}
we obtain
\begin{equation}\label{Cirunsol}
   \vec{C}\big(\alpha_s(\mu)\big) 
   \approx  \exp\left[ - \frac{\bm{\gamma}_0^T}{2\beta_0}\,\ln\frac{\alpha_s(\mu)}{\alpha_s(M)} \right] 
    \vec{C}\big(\alpha_s(M)\big) \,.
\end{equation}
The matrix exponential can easily be evaluated in {\sc Mathematica} (problem~4.2). For the simplest case of a single operator ${\mathcal O}$, we find
\begin{equation}\label{eq123}
   C\big(\alpha_s(\mu)\big) 
   \approx \left( \frac{\alpha_s(\mu)}{\alpha_s(Q)} \right)^{-\gamma_0/2\beta_0} C\big(\alpha_s(M)\big) \,.
\end{equation}
This solution is analogous to that for the running mass in (\ref{mqrunresu}).

We can use the solution (\ref{Cirunsol}) to obtain the effective Lagrangian (\ref{EFTmu}) at the low-energy scale $\mu=m$, which is characteristic for the mass scale of the low-energy effective theory. At this scale, the matrix elements of the local operators ${\mathcal O}_i(\mu)$ evaluated between physical states can be calculated in fixed-order perturbation theory, since they are free of large logarithms. All potentially large logarithmic corrections are contained in the Wilson coefficients $C_i\big(\alpha_s(m)\big)$. To see in detail how the large logarithms are resummed, we can substitute from (\ref{eq93}) the relation
\begin{equation}
   \frac{\alpha_s(m)}{\alpha_s(M)} \approx \left[ 1-\beta_0\,\frac{\alpha_s(M)}{4\pi}\,\ln\frac{M^2}{m^2} \right]^{-1}
\end{equation}
for the ratio of coupling constants in (\ref{eq123}), focussing for simplicity on the case of a single operator. This yields
\begin{equation}
\begin{aligned}
   C(m) &\approx  \left( 1 - \beta_0\,\frac{\alpha_s(M)}{4\pi}\,\ln\frac{M^2}{m^2} \right)^{\gamma_0/2\beta_0} C(M) \\
   &= \left[ 1 - \frac{\gamma_0}{2}\,\frac{\alpha_s(M)}{4\pi}\,\ln\frac{M^2}{m^2}
    + \frac{\gamma_0(\gamma_0-2\beta_0)}{8} \left( \frac{\alpha_s(M)}{4\pi}\,\ln\frac{M^2}{m^2} \right)^2
    + \dots \right] C(M) \,.
\end{aligned}
\end{equation}
For $\frac{\alpha_s(M)}{4\pi}\,\ln\frac{M^2}{m^2}=O(1)$ each term in the series contributes at the same order, and resummation is necessary in order to obtain a reliable result.

\newpage
\section{\boldmath One last remark concerning the running QCD coupling}
\label{sec:5.7}

In order for the above expressions for the Wilson coefficients to make sense, we need to make sure that our formula for $\alpha_s(\mu)$ can be reliably evaluated at any value of $\mu$ in the perturbative regime ($\mu\gg\Lambda_{\rm QCD}$). At leading order we found in (\ref{asrun})
\begin{equation}
   \alpha_s(\mu) \approx \frac{\alpha_s(Q)}{1+\beta_0\,\frac{\alpha_s(Q)}{4\pi}\,\ln\frac{\mu^2}{Q^2}} \,.
\end{equation}
One might worry what happens if the logarithm in the denominator becomes large. In other words, we need to demonstrate that higher-order corrections in the $\beta$-function do not spoil this formula by introducing additional large logarithms. To see that this does indeed not happen, we keep the next term in the perturbative series for the $\beta$-function in (\ref{betafunc}) and study its effect on the solution for the running coupling, which is obtained from (\ref{betadef}). Separating variables, we obtain
\begin{equation}
   - \frac{d\alpha_s}{\alpha_s^2}\,
    \frac{1}{1 + \frac{\beta_1}{\beta_0}\,\frac{\alpha_s}{4\pi} + \dots}
   = \frac{\beta_0}{4\pi}\,d\ln\mu^2 \,.
\end{equation}
Note that the right-hand side is the single source of logarithms, while no logarithms appear on the left-hand side. As long as we are in the perturbative regime where $\frac{\alpha_s}{4\pi}\ll 1$, we can expand the left-hand side in a perturbative series and obtain, at next-to-leading order,
\begin{equation}
   - \frac{d\alpha_s}{\alpha_s^2} \left( 1 - \frac{\beta_1}{\beta_0}\,\frac{\alpha_s}{4\pi} + \dots \right)
   = \frac{\beta_0}{4\pi}\,d\ln\mu^2 \,.
\end{equation}
Integrating this equation gives
\begin{equation}
   \frac{1}{\alpha_s(\mu)} - \frac{1}{\alpha_s(Q)}
    + \frac{\beta_1}{4\pi\beta_0}\,\ln\frac{\alpha_s(\mu)}{\alpha_s(Q)}
    + O\bigg(\frac{\alpha_s(\mu)-\alpha_s(Q)}{16\pi^2}\bigg)
   = \frac{\beta_0}{4\pi}\,\ln\frac{\mu^2}{Q^2} \,.
\end{equation}
Multiplying both side with $\alpha_s(Q)$ gives
\begin{equation}
   \frac{\alpha_s(Q)}{\alpha_s(\mu)}
    - \frac{\beta_1}{\beta_0}\,\frac{\alpha_s(Q)}{4\pi}\,\ln\frac{\alpha_s(Q)}{\alpha_s(\mu)}
    + O\bigg(\frac{\alpha_s(Q)}{4\pi}\,\frac{\alpha_s(\mu)-\alpha_s(Q)}{4\pi}\bigg)
   = 1 + \beta_0\,\frac{\alpha_s(Q)}{4\pi}\,\ln\frac{\mu^2}{Q^2} \,.
\end{equation}
Once again, the only potentially large logarithm is the one on the right-hand side. We can now insert, in an iterative way, the leading-order solution for $\alpha_s(Q)/\alpha_s(\mu)$ in the second term on the left-hand side to obtain
\begin{equation}
   \frac{\alpha_s(Q)}{\alpha_s(\mu)}
   = 1 + \beta_0\,\frac{\alpha_s(Q)}{4\pi}\,\ln\frac{\mu^2}{Q^2}
    + \frac{\beta_1}{\beta_0}\,\frac{\alpha_s(Q)}{4\pi}\,
    \ln\bigg( 1 + \beta_0\,\frac{\alpha_s(Q)}{4\pi}\,\ln\frac{\mu^2}{Q^2} \bigg)
    + \dots \,.
\end{equation}
Even in the ``large-log region'', where $\frac{\alpha_s(Q)}{4\pi}\,\ln\frac{\mu^2}{Q^2}=O(1)$ or larger, the correction proportional to $\beta_1$ (the two-loop coefficient of the $\beta$-function) is suppressed by at least $\frac{\alpha_s(Q)}{4\pi}\ll 1$ relative to the leading term. The leading-order formula for $\alpha_s(\mu)$ is thus a decent approximation for all values $\mu\gg\Lambda_{\rm QCD}$.

\section{Homework problems}

\begin{enumerate}
\item[4.1]
Derive relation (\ref{moremagic}), and clarify the origin of the minus sign between this equation and~(\ref{gammamagic}).
\item[4.2]
The effective weak Lagrangian for the nonleptonic decay $\bar B^0\to\pi^+ D_s^-$ of the neutral $B$ meson contains two dimension-6 four-fermion operators, which differ in their color structure. Specifically, one finds (here $i,j$ are color indices)
\begin{equation}\label{pp12}
   {\mathcal L}_{\rm eff} = - \frac{4G_F}{\sqrt2}\,V_{cs}^*\,V_{ub}
    \left[ C_1(\mu)\,\bar s_L^j\gamma_\mu c_L^j\,\bar u_L^i\gamma^\mu b_L^i
    + C_2(\mu)\,\bar s_L^i\gamma_\mu c_L^j\,\bar u_L^j\gamma^\mu b_L^i \right] ,
\end{equation}
where $C_1=1+O(\alpha_s)$ and $C_2=O(\alpha_s)$ follow from tree-level matching of the $W$-boson exchange diagram onto the effective theory. Using a Fierz rearrangement, the second operator above can also be 
written as $\bar u_L^j\gamma_\mu c_L^j\,\bar s_L^i\gamma^\mu b_L^i$. Note
also that
\begin{equation}
   \bar s_L\gamma_\mu t_a c_L\,\bar u_L\gamma^\mu t_a b_L
   = \frac12\,\bar s_L^i\gamma_\mu c_L^j\,\bar u_L^j\gamma^\mu b_L^i
   - \frac{1}{2N_c}\,\bar s_L\gamma_\mu c_L\,\bar u_L\gamma^\mu b_L
\end{equation}
by virtue of a color Fierz identity, where $t_a$ are the generators of color $SU(N_c)$. By computing the UV divergences of the two operators in (\ref{pp12}) at one-loop order (including the effects of wave-function renormalization), show that the anomalous-dimension matrix for the two operators takes the form
\[
   \bm{\gamma} = \frac{\alpha_s}{4\pi} \left(
    \begin{array}{cc}
     - \frac{6}{N_c} & 6 \\
     6 & - \frac{6}{N_c} \\
    \end{array} \right) + O(\alpha_s^2) \,.
\]
Given this result, work out the explicit form of the leading-order solution to the RG equation (\ref{eq118}), which has been given in (\ref{Cirunsol}).
\end{enumerate}

\end{document}